\RequirePackage[2020-02-02]{latexrelease}
\documentclass[nopreprintnumbers,10pt,a4paper,notitlepage,nofootinbib,longbibliography,aps,prd,amsfonts,amssymb,amsmath]{revtex4-2}

\usepackage{hyperref}
\usepackage{graphicx}
\usepackage{color}

\begin{document}

\title{Non-relativistic stellar structure in higher-curvature gravity: systematic construction of solutions to the modified Lane--Emden equations}

\author{Shinpei Tonosaki}\email[]{tonosaki(a)tap.st.hirosaki-u.ac.jp}
\affiliation{
Graduate School of Science and Technology, Hirosaki University,
Hirosaki, Aomori 036-8561, Japan
}
\author{Tomoya Tachinami}\email[Corresponding author. ]{tachinami(a)tap.st.hirosaki-u.ac.jp}
\affiliation{
Graduate School of Science and Technology, Hirosaki University,
Hirosaki, Aomori 036-8561, Japan
}
\author{Yuuiti Sendouda}\email[]{sendouda(a)hirosaki-u.ac.jp}
\affiliation{
Graduate School of Science and Technology, Hirosaki University,
Hirosaki, Aomori 036-8561, Japan
}

\date{\today}

\begin{abstract}
We study the structure of static spherical stars made up of a non-relativistic polytropic fluid in linearized higher-curvature theories of gravity (HCG).
We first formulate the modified Lane--Emden (LE) equation for the stellar profile function in a gauge-invariant manner, finding it boils down to a sixth order differential equation in the generic case of HCG, while it reduces to a fourth order equation in two special cases, reflecting the number of additional massive gravitons arising in each theory.
Moreover, the existence of massive gravitons renders the nature of the boundary-value problem unlike the standard LE:
some of the boundary conditions can no longer be formulated in terms of physical conditions at the stellar center alone, but some demands at the stellar surface necessarily come into play.
We present a practical scheme for constructing solutions to such a problem and demonstrate how it works in the cases of the polytropic index $ n = 0 $ and $ 1 $, where analytical solutions to the modified LE equations exist.
As physical outcomes, we clarify how the stellar radius, mass, and Yukawa charges depend on the theory parameters and how these observables are mutually related.
Reasonable upper bounds on the Weyl-squared correction are obtained.
\end{abstract}

\maketitle

\section{Introduction}

Motivations for considering higher-curvature theories of gravity (HCG), whose action is generically of the form
\begin{equation}
S_\mathrm g
= \int\!\mathrm d^4x\,\sqrt{-g}\,f(R^\mu{}_{\nu\rho\sigma}, g_{\mu\nu})
\label{eq:fRiemann}
\end{equation}
with $ f $ being any non-linear scalar function of the Riemann tensor $ R^\mu{}_{\nu\rho\sigma} $ and metric $ g_{\mu\nu} $\,, come from several aspects of gravitation and even date back to the earliest days in the history of relativistic theories of gravity.
Viewed as quantum corrections to General Relativity (GR), various possibilities of $ f $ have been so far argued, e.g., for a resolution of cosmological singularities \cite{Starobinsky:1980te}, a cure for the ultraviolet divergence \cite{Stelle:1976gc}, and so on.
A rather modern view is to regard it as a low-energy effective theory descending from as-yet-unrevealed high-energy physics like superstring theories.
See, e.g., \cite{Clifton:2011jh,Belenchia:2016bvb} for general reviews on HCG.

Classical dynamics of HCG can be characterized in terms of the inherent dynamical gravitational degrees of freedom (dofs), which most clearly manifest in the linear approximation.
When deviation from the flat space-time, $ h_{\mu\nu} \equiv g_{\mu\nu} - \eta_{\mu\nu} $\,, is small, the generic action \eqref{eq:fRiemann} can be approximated by lowest-order terms in curvature.
In the linear approximation, only terms up to quadratic order are relevant, and such terms can be organized as
\begin{equation}
S_\mathrm g
= \frac{1}{16\pi\,G}\,\int\!\mathrm d^4x\,\sqrt{-g}\,
  \left(
   R
   - \alpha\,C_{\mu\nu\rho\sigma}\,C^{\mu\nu\rho\sigma}
   + \beta\,R^2
   + \mathcal O(R^\mu{}_{\nu\rho\sigma})^3
  \right)\,,
\label{eq:action}
\end{equation}
where $ G $ is the Newton constant, $ R $ is the Ricci scalar, $ C_{\mu\nu\rho\sigma} $ is the Weyl curvature tensor, the coefficients $ \alpha $ and $ \beta $ have the units of length squared, and we have made a reasonable assumption on $ f $ that GR should recover in the small curvature limit.
Note that the above action is fully general up to quadratic order owing to the fact that the Gauss--Bonnet combination is topological in four dimensions.
As revealed by Stelle \cite{Stelle:1977ry}, the particle content of the above full quadratic curvature gravity (QCG) consists of, on top of the zero-mass spin-$ 2 $ graviton, a spin-$ 2 $ graviton with ``mass'' $ m_2 = 1/\sqrt{2\alpha} $ and a spin-$ 0 $ graviton with ``mass'' $ m_0 = 1/\sqrt{6\beta} $.\footnote{To be precise, $ m_2 $ and $ m_0 $ have the dimension of inverse length.}
Also shown by him is that the massive spin-$ 2 $ is a ghost with negative kinetic energy and provides a repulsive Yukawa force, whereas the massive spin-$ 0 $ is attractive.
Afterwards, analogous analyses on the dynamical dofs of QCG in non-flat background geometries were carried out in the case of (anti-)de Sitter space \cite{Tekin:2016vli} and even any Einstein manifolds \cite{Niiyama:2019fvf}.
On the other hand, a background-independent hamiltonian analysis of the full $ f(\text{Riemann}) $ gravity was presented in \cite{Deruelle:2009zk}.
For a recent review on QCG, see \cite{Salvio:2018crh}.

In order to seek for possible observable signatures of HCG and their consequences, a full study on the polarizations of gravitational waves on the flat background has been conducted recently by the present authors in \cite{Tachinami:2021jnf}.
As remarked, what is significantly useful in studies of linearized HCG is its equivalence to a multi-graviton theory.
In \cite{Tachinami:2021jnf}, we have employed a gauge-invariant formalism and identified all the independent helicity variables that propagate in vacuum.
Future developments in gravitational-wave observations along this direction should help examine the viability of HCG.

In this paper, our interest is drawn to non-vacuum situations in which gravity is sourced by non-relativistic fluid matter.
As the characteristic of the matter source, we adopt the polytropic relation $ P = K\,\rho^{1+\frac{1}{n}} $ between the pressure $ P $ and mass density $ \rho $ as the equation of state (eos) and consider static spherically symmetric configuration.
In the newtonian limit, the matter equation of motion (eom) in such a setup settles down to the hydrostatic equilibrium condition in a familiar form.
In GR, this condition can be cast into the so-called Lane--Emden (LE) equation.
Solutions to the LE equation offer simple yet reasonable stellar structure models.
Moreover, resultant relations between the stellar characteristic quantities such as the radius, mass, and central density can be used as scrutinies of adopted laws of physics such as the equation of state of the constituting matter.

Naturally and importantly, stellar structure is quite sensitive to any modifications of gravity.
There have been various attempts to probe gravity at work in the stellar interior, including general polytropic stars \cite{Chen:2001a,10.1143/PTP.106.63,Capozziello:2011nr,Farinelli:2013pza,Saito:2015fza,Wojnar:2018hal,Sergyeyev:2019aul,Fabris:2021qkp,Sharif:2022nrm,Chowdhury:2022qvv}, nuclear-burning stars \cite{Sakstein:2015zoa,Sakstein:2016lyj,Andre:2017usv,Cermeno:2018qed}, white dwarfs \cite{Banerjee:2017uwz,Kalita:2022trq}, and neutron stars \cite{Cooney:2009rr,Arapoglu:2010rz,Deliduman:2011nw,Bonanno:2021zoy}.
We refer the reader to an extensive review paper on these subjects by Olmo \textit{et al.} \cite{Olmo:2019flu}, while we feel necessity of supplementing the following remarks greatly relevant to the present research;
To our best knowledge, an analogue of the LE equation in generic QCG was derived by Chen and Shao in a little-noticed paper in 2001 \cite{Chen:2001a}.
They incorporated Yukawa potentials of the extra massive gravitons into the hydrostatic equilibrium condition and derived a sixth order differential equation.
They even obtained analytical solutions for the polytropic indices $ n = 0 $ and $ 1 $, although they seem incorrectly presented for some reason and their analysis on the observables was limited to the perturbative changes in the stellar radius.
Also, along the same line, a simpler ``$ R+R^2 $'' version of the LE equation, which is a fourth order equation, was derived in Chen \textit{et al.} \cite{10.1143/PTP.106.63} in 2001 and analytical solutions for $ n = 0 $ and $ 1 $ were found.
Unfortunately, these 2001 papers seem to have drawn little attention until very recently (even from the present authors), and they are missing in the review \cite{Olmo:2019flu}.
Instead, people have followed an equivalent but less tractable formulation presented much later in the $ f(R) $ context \cite{Capozziello:2011nr}.

The goal of our present study will be to perform diagnostic of HCG by means of astrophysical observations of stellar structure.
In doing so, we have to deal with higher-order LE equations in HCG.
Since the modified LE equation is, in the maximal case, a sixth-order differential equation, it requires six boundary conditions.
Furthermore, as experienced in past studies \cite{Chen:2001a,10.1143/PTP.106.63,Capozziello:2011nr}, an essential discrepancy from the standard, second-order case we are to face is that the problem is no longer formulated as a simple initial value problem.
We first clarify the origin of this property in a gauge-invariant formalism.
Then we rigorously derive the boundary conditions for the higher-order LE equations from the hydrostatic equilibrium condition.
In the meantime, we will highlight the difference of our treatment from the past ones.
Finally, we present a systematic way of constructing solutions compatible with our boundary conditions and demonstrate how it works for the cases of $ n = 0 $ and $ 1 $, taking advantage of the existence of analytical solutions.
Though out of scope of this paper, our scheme is applicable to any polytropic index $ n $ for which numerical integration is inevitable.

The organization of this paper is as follows.
In Sec.~\ref{sec:le}, we consider hydrostatic equilibrium in a non-relativistic static star in linearized HCG.
We employ a gauge-invariant formalism and derive basic equations for gravitational potentials.
Then adopting the polytropic relation as the eos, we derive the modified LE equation.
In Sec.~\ref{sec:bc}, we derive the boundary conditions from the behavior of the gravitational potentials at the stellar center via the hydrostatic equilibrium condition.
In Sec.~\ref{sec:sol}, we demonstrate a scheme for the construction of the solutions to the modified LE equations.
In the meantime, we discuss semi-observational bounds on the QCG modifications to GR.
Finally we summarize the present work and give some conclusions in Sec.~\ref{sec:concl}.

Throughout the paper, we will work with natural units with $ c = \hbar = 1 $.
Greek indices of tensors such as $ \mu,\nu,\cdots $ are of space-time while Latin ones such as $ i,j,\cdots $ are spatial.
We adopt the mostly-positive sign convention for the metric, so the Minkowski metric in the Cartesian coordinates is $ \eta_{\mu\nu} = \mathrm{diag}(-1,1,1,1) $.
$ \square \equiv \eta^{\mu\nu}\,\partial_\mu \partial_\nu $ and $ \triangle \equiv \delta^{ij}\,\partial_i \partial_j $ are the d'Alembert and Laplace operators in the flat background.
The Riemann tensor is defined as $ R^\mu{}_{\nu\rho\sigma} = \partial_\rho \Gamma^\mu{}_{\nu\sigma} - \cdots $\,.

\section{\label{sec:le}Hydrostatic equilibrium and the modified LE equations in HCG}

In this section, we formulate the hydrostatic equilibrium condition in linearized HCG in terms of gauge-invariant variables.
Adopting a polytropic equation of state, together with the assumption of staticity and spherical symmetry, we reproduce the single higher-order differential equations for the stellar profile function derived in \cite{Chen:2001a,10.1143/PTP.106.63}.
The derived equations can be viewed as extensions of the standard LE equation in GR.

\subsection{Equations of motion}

In order to derive linear equations of motion, we begin with calculating the second-order perturbation of the gravitational action \eqref{eq:action} plus a minimally coupled conservative matter.
As the matter source, we adopt a perfect fluid with energy density $ \epsilon $ and pressure $ P $ coupled with scalar-type gravitational perturbations.
The scalar sector of the second-order perturbative action is then given by
\begin{equation}
{}^{(2)}S
= \frac{1}{16\pi\,G}\,\int\!\mathrm d^4x\,\left[
   -6 \Phi\,\square \Phi - 4 \Phi\,\triangle (\Psi-\Phi)
   - \frac{4}{3}\,\alpha\,[\triangle (\Psi-\Phi)]^2
   + \beta\,\left({}^{(1)}R\right)^2
  \right]
  + \int\!\mathrm d^4x\,\left[
     - \Psi\,\epsilon
     + 3 \Phi\,P
    \right]\,,
\label{eq:action_s}
\end{equation}
where $ \Psi $ and $ \Phi $ are gauge-invariant linear scalar perturbations and
\begin{equation}
{}^{(1)}R
= -6 \square\Phi - 2 \triangle (\Psi-\Phi)
\end{equation}
is the Ricci scalar at the linear order.
See Appendix~\ref{sec:gauge} for the definition of the gauge-invariant variables and \ref{sec:pert} for the derivation of the action.
Varying the action with respect to $ \Psi $ and $ \Phi $, we obtain the field equations
\begin{equation}
\begin{aligned}
\triangle \left[
 -2 \Phi
 - \frac{4}{3} \alpha\,\triangle (\Psi-\Phi)
 - 2 \beta\,{}^{(1)}R
\right]
&
= 8 \pi\,G\,\epsilon\,, \\
{}^{(1)}R
+ 2 \triangle \Phi
+ \frac{4}{3}\,\alpha\,\triangle^2 (\Psi - \Phi)
+ 2 \beta\,(-3 \square + \triangle) {}^{(1)}R
&
= -24 \pi\,G\,P\,.
\end{aligned}
\label{eq:eom}
\end{equation}

In the static case, the coupled equations \eqref{eq:eom} can be reduced and reorganized in a neat way.
To see this, we introduce an alternative set of gauge-invariant variables
\begin{equation}
\begin{aligned}
\Psi_2
&
\equiv
  \frac{1}{2}\,\Psi - \frac{1}{2}\,\Phi\,, \\
\Psi_0
&
\equiv
  -\Psi - 2 \Phi\,,
\end{aligned}
\label{eq:Psi_s}
\end{equation}
where, as will be clarified shortly, the subscripts refer to the spin $ s $ of the massive gravitons.
For later convenience we denote the inversion of the above as
\begin{equation}
\begin{aligned}
\Psi
&
= -\alpha_2\,\Psi_2  - \alpha_0\,\Psi_0\,, \\
\Phi
&
= \beta_2\,\Psi_2 + \beta_0\,\Psi_0
\end{aligned}
\end{equation}
with
\begin{equation}
\alpha_2
= -\frac{4}{3}\,,
\quad
\alpha_0
= \frac{1}{3}\,,
\quad
\beta_2
= -\frac{2}{3}\,,
\quad
\beta_0
= -\frac{1}{3}\,.
\label{eq:conv}
\end{equation}
Note that these coefficients are normalized so that $ \alpha_2 + \alpha_0 = -1 $ and $ \beta_2 + \beta_0 = -1 $.
By taking linear combinations of \eqref{eq:eom} after reducing $ \square $ to $ \triangle $, we obtain the decoupled equations for the gauge-invariant variables $ \Psi_s $ as
\begin{equation}
\begin{aligned}
\triangle \Psi_2
- 2 \alpha\,\triangle^2 \Psi_2
&
= 4\pi\,G\,\left(\epsilon + \frac{3}{2}\,P\right)\,, \\
\triangle \Psi_0 - 6 \beta\,\triangle^2 \Psi_0
&
= 4\pi\,G\,\left(\epsilon - 3 P\right)\,.
\end{aligned}
\end{equation}
One may already notice the similarity of the above two equations.
Furthermore, with the non-relativistic approximation $ |P| \ll \epsilon \simeq \rho $ with $ \rho $ being the rest mass density of the fluid, these reduce to exactly the same form
\begin{equation}
\left(\triangle - m_s^2\right)\,\triangle \Psi_s
= -4\pi\,G\,\rho\,m_s^2
\label{eq:eom_Psi_s}
\end{equation}
with $ s = 2,0 $, where we have introduced the ``mass'' parameters $ m_2^2 \equiv 1/(2\alpha) $ and $ m_0^2 \equiv 1/(6\beta) $.

The key property inherent in these fourth-order equations is that the gauge-invariant potentials $ \Psi_s $ can be expressed as the difference of two variables
\begin{equation}
\Psi_s
= \phi_s - \psi_s\,,
\end{equation}
where $ \phi_s $ and $ \psi_s $ are required to satisfy the following Poisson and Helmholtz equations,
\begin{align}
\triangle \phi_s
&
= 4\pi\,G\,\rho\,,
\label{eq:P} \\
\left(\triangle - m_s^2\right)\,\psi_s
&
= 4\pi\,G\,\rho\,,
\label{eq:H}
\end{align}
respectively.
An immediate implication of this decomposition being possible is that $ \phi_s $ is the massless graviton in GR while $ \psi_s $ originates from the extra massive dof with spin $ s $ arising in HCG.
Indeed, these Helmholtz equations are reminiscent of the Klein--Gordon equations satisfied by the helicity-$ 0 $ dofs with the same masses identified in \cite{Tachinami:2021jnf}.
Note, however, that there is a certain difference in the current definition of the variables from the vacuum case.
See also \cite{Chen:2001a,10.1143/PTP.106.63} for an analogous discussion for decomposing a higher-order equation into second-order equations in the presence of matter.
Since the Poisson equations \eqref{eq:P} for both $ s $ are identical, and so are their solutions, we may omit the subscript $ s $ for $ \phi_s $: $ \phi \equiv \phi_2 = \phi_0 $.
Then the original gauge-invariant variables can be represented as
\begin{align}
\Psi
&
= \phi + \alpha_2\,\psi_2 + \alpha_0\,\psi_0\,,
\label{eq:Psi} \\
\Phi
&
= -\phi + \beta_2\,\psi_2 + \beta_0\,\psi_0
\label{eq:Phi}
\end{align}
with the coefficients given by \eqref{eq:conv}.

It may be worth mentioning here the ``massive'' limit with $ m_s^2 \to \infty $.
Then, only terms in proportion to $ m_s^2 $ in the fourth-order equation \eqref{eq:eom_Psi_s} remain, reducing to a conventional Poisson equation for $ \Psi_s $\,.
Also, Eq.~\eqref{eq:H} in this limit imposes $ \psi_s = 0 $, and we have $ \Psi_s = \phi $.
Another interesting limit is the massless limit, $ m_s^2 \to 0 $, where one expects $ \psi_s \to \phi $.
However, a caution to be given here is that a vanishingly small value of the mass parameter $ m_s^2 $ corresponds to a huge absolute value of the expansion coefficient in the action \eqref{eq:action} ($ \alpha $ or $ \beta $), for which the validity of the small-curvature approximation may be questioned.

\subsection{Gravitational potentials}

Our next task is to express the gravitational potentials in a compatible way with adequate boundary conditions;
On the one hand, asymptotic flatness requires $ \lim_{|\vec x| \to \infty} \Psi = \lim_{|\vec x| \to \infty} \Phi = 0 $\,, so the boundary conditions to be imposed on $ \Psi_s $ are $ \lim_{|\vec x| \to \infty} \Psi_2 = \lim_{|\vec x| \to \infty} \Psi_0 = 0 $.
Also, all these perturbative variables must be bounded in the whole spatial domain.
The same should also hold for $ \phi $ and $ \psi_s $\,.

Hereafter we specialize to a spherically symmetric configuration, where every quantity becomes a function of the radial coordinate $ r $ and the Laplace operator reduces to $ \triangle = \frac{1}{r^2}\,\frac{\mathrm d}{\mathrm dr}\left(r^2\,\frac{\mathrm d}{\mathrm dr}\right) $.
We shall impose regularity at the stellar center $ r = 0 $ and flatness at $ r \to \infty $.
The solution to the Poisson equation \eqref{eq:P} satisfying these demands is the conventional newtonian potential:
\begin{equation}
\phi_s
= \phi
= -G\,\int_r^\infty\!\mathrm dr'\,\frac{m(r')}{r'^2}
\label{eq:phi}
\end{equation}
with the enclosed mass $ m(r) $ being
\begin{equation}
m(r)
\equiv
  4\pi\,\int_0^r\!\mathrm dr'\,r'^2\,\rho(r')\,.
\label{eq:mass}
\end{equation}
If the matter is confined within a finite radius $ R $ like a star, $ \rho(r \geq R) = 0 $, then the enclosed mass is constant outside the stellar radius, $ m(r \geq R) = m(R) \equiv M $\,, so the exterior potential is
\begin{equation}
\phi(r \geq R)
= -\frac{G\,M}{r}\,.
\label{eq:phi_ext}
\end{equation}
The total mass $ M $ can be measured at remote distances using Kepler's law, for instance.

On the other hand, we shall treat the massive modes with a little more care.
The general solution to the Helmholtz equation \eqref{eq:H} with two integration constants, $ C_1 $ and $ C_2 $\,, is
\begin{equation}
\psi_s
= -\frac{G}{r}\,
  \left(
   \sigma_s(r)\,\cosh m_s\,r
   - \chi_s(r)\,\sinh m_s\,r
   + C_1\,\sinh m_s\,r
   + C_2\,\cosh m_s\,r
  \right)
\end{equation}
with the two functions $ \sigma_s $ and $ \chi_s $ with the dimensions of mass being
\begin{equation}
\begin{aligned}
\sigma_s(r)
&
\equiv
  \frac{4\pi}{m_s}\,\int_0^r\!\mathrm dr'\,r'\,\rho(r')\,\sinh m_s\,r'\,, \\
\chi_s(r)
&
\equiv
  \frac{4\pi}{m_s}\,\int_0^r\!\mathrm dr'\,r'\,\rho(r')\,\cosh m_s\,r'\,.
\end{aligned}
\label{eq:charge}
\end{equation}
For a star with the finite radius $ R $, likewise, these quantities have a constant value outside the star:
\begin{equation}
\sigma_s(r \geq R)
= \sigma_s(R)
\equiv
  \Sigma_s\,,
\quad
\chi_s(r \geq R)
= \chi_s(R)\
\equiv
  X_s\,.
\end{equation}
In this case, regularity at $ r = 0 $ requires $ C_2 = 0 $, while asymptotic flatness fixes the other constant as
\begin{equation}
C_1
= X_s - \Sigma_s
= \frac{4\pi}{m_s}\,\int_0^R\!\mathrm dr\,r\,\rho(r)\,\mathrm e^{-m_sr}
\equiv
  I_s\,.
\end{equation}
Thus, the spin-dependent massive potential is determined as
\begin{equation}
\psi_s
= -\frac{G}{r}\,\left(
   \sigma_s(r)\,\cosh m_s\,r + (I_s - \chi_s(r))\,\sinh m_s\,r
  \right)\,.
\label{eq:psi}
\end{equation}
Outside the star, it reduces to a Yukawa-type potential characterized by the graviton mass $ m_s $ and the total ``charge'' $ \Sigma_s $ as
\begin{equation}
\psi_s(r \geq R)
= -\frac{G\,\Sigma_s\,\mathrm e^{-m_s r}}{r}\,.
\label{eq:psi_ext}
\end{equation}
As a consequence, the total gravitational potential $ \Psi $ within the distance $ m_s^{-1} $ from the star surface acquires Yukawa-type modifications.

Note that the enclosed ``charges'' $ \sigma_s $ and $ \chi_s $ are both positive-semidefinite for positive mass density $ \rho $ as is the case for the enclosed mass $ m $.
They are divergent in the massive limit, $ m_s \to \infty $, but the extra potential $ \psi_s $ converges to $ 0 $ in this limit.
On the other hand, in the massless limit, $ m_s \to 0 $, the enclosed charge $ \sigma_s $ tends to the same value as the enclosed mass as $ \sigma_s = m + \mathcal O(m_s^2) $, while $ \chi_s $ is divergent as $ \chi_s = \mathcal O(m_s^{-1}) $, and $ \psi_s $ tends to the conventional potential $ \phi $ as expected.

Since these gravitational potentials have been so constructed to satisfy Eq.~\eqref{eq:P} or \eqref{eq:H}, their derivatives are necessarily continuous up to second order everywhere.
On the other hand, the third derivatives involve $ \mathrm d\rho/\mathrm dr $ and therefore are discontinuous at the stellar surface.

\subsection{Hydrostatic equilibrium: Sixth-order master equation}

For the time being, we assume the graviton mass parameters $ m_s^2 $'s are both bounded, leaving discussions of the limiting cases to the next subsection, in which either or both of $ m_s^2 $'s are taken to infinity.

Given the gauge-invariant potential $ \Psi $, the hydrostatic equilibrium condition in the newtonian gauge reads
\begin{equation}
\frac{1}{\rho}\,\frac{\mathrm dP}{\mathrm dr}
= -\frac{\mathrm d\Psi}{\mathrm dr}\,,
\label{eq:hydro}
\end{equation}
where the (outward) radial acceleration $ -\frac{\mathrm d\Psi}{\mathrm dr} $ is obtained by differentiating \eqref{eq:Psi} as
\begin{equation}
-\frac{\mathrm d\Psi}{\mathrm dr}
= -\frac{G\,m}{r^2}
  - \alpha_2\,\frac{\mathrm d\psi_2}{\mathrm dr}
  - \alpha_0\,\frac{\mathrm d\psi_0}{\mathrm dr}\,.
\label{eq:accel}
\end{equation}
One can deduce from the above expression that the sign of the coefficient $ \alpha_s $ is crucial in determining the direction of the force exerted by the massive graviton of spin $ s $, i.e., attractive or repulsive.
Due to the reverse signs of $ \alpha_2 = -4/3 $ and $ \alpha_0 = 1/3 $\,, the two massive gravitons work in opposite ways.

By supplying an equation of state $ P = P(\rho) $\,, the hydrostatic equilibrium condition \eqref{eq:hydro} reduces to an equation for the single function $ \rho $.
At this stage, however, \eqref{eq:hydro} is an integro-differential equation for $ \rho $ since the functions $ m $ and $ \psi_s $ on the right-hand side involve radial integrations.
In other words, in order to obtain an equivalent differential equation, one has to extract $ \rho $ from $ m $ and $ \psi_s $.
Let us first follow the standard procedure in GR, i.e., operate $ \frac{1}{r^2}\,\frac{\mathrm d}{\mathrm dr} r^2 $ on the both sides:
\begin{equation}
\frac{1}{r^2}\,
\frac{\mathrm d}{\mathrm dr} \left(\frac{r^2}{\rho}\,\frac{\mathrm dP}{\mathrm dr}\right)
= -4 \pi\,G\,\rho
  - \alpha_2\,\triangle \psi_2
  - \alpha_0\,\triangle \psi_0\,.
\end{equation}
Since we have assumed both $ m_s^2 $'s are bounded, we can safely use the Helmholtz equations \eqref{eq:H} for $ \psi_2 $ and $ \psi_0 $ to obtain
\begin{equation}
\frac{1}{r^2}\,
\frac{\mathrm d}{\mathrm dr} \left(\frac{r^2}{\rho}\,\frac{\mathrm dP}{\mathrm dr}\right)
= -\alpha_2\,m_2^2\,\psi_2
  - \alpha_0\,m_0^2\,\psi_0\,.
\label{eq:hydro_psi}
\end{equation}
Here, a crucial difference from the case of GR is that this is still an integro-differential equation since $ \psi_s $ involves integrals of $ \rho $ in a non-trivial manner.
Another striking difference from GR is that the standard source term $ -4 \pi G\,\rho $ has been cancelled out.
These would appear to imply a thorough change in the structure of the governing equation compared to GR.
We will see, however, one can take an appropriate limit of the above equation to find the correct expression in GR.

Hereafter we adopt the polytropic relation $ P = K\,\rho^{1+\frac{1}{n}} $ as the eos for the stellar matter, where $ K $ is the normalization constant and $ n $ is the constant called the polytropic index.
Introduce a length scale
\begin{equation}
\ell
\equiv
  \sqrt{\frac{(n+1)\,P_\mathrm c}{4\pi\,G\,\rho_c^2}}\,,
\label{eq:ell}
\end{equation}
where the quantities with the subscript ``c'' denote the values at the stellar center, such as $ \rho_\mathrm c \equiv \rho(r=0) $ and $ P_\mathrm c \equiv P(r=0) = K\,\rho_\mathrm c^{1+\frac{1}{n}} $.
The non-dimensional radial coordinate $ \xi $ and graviton mass parameters $ \mu_s $ are defined by
\begin{equation}
\xi
\equiv
  \frac{r}{\ell}\,,
\quad
\mu_0
\equiv
  m_0\,\ell\,,
\quad
\mu_2
\equiv
  m_2\,\ell\,.
\end{equation}
The stellar mass density and pressure are conveniently replaced by the non-dimensional profile function $ \theta(\xi) $ as
\begin{equation}
\rho
= \rho_\mathrm c\,\theta^n\,,
\quad
P
= P_\mathrm c\,\theta^{n+1}\,.
\end{equation}
By definition, $ \theta $ is normalized to unity at the center, $ \theta_\mathrm c \equiv \theta(\xi=0) = 1 $.
The value of $ \xi $ at the first positive zero of $ \theta $ is denoted as $ \xi_R $ and is related to the stellar radius as $ R = \xi_R\,\ell $.
The laplacian operator is non-dimensionalized as $ \triangle_\xi \equiv \frac{1}{\xi^2}\,\frac{\mathrm d}{\mathrm d\xi} \left(\xi^2\,\frac{\mathrm d}{\mathrm d\xi}\right) = \ell^2\,\triangle $, and the hydrostatic equilibrium condition \eqref{eq:hydro_psi} for the polytropic eos reads
\begin{equation}
\triangle_\xi \theta
= -\frac{\alpha_0\,\mu_0^2\,\psi_0 + \alpha_2\,\mu_2^2\,\psi_2}{4\pi\,G\,\rho_\mathrm c\,\ell^2}\,.
\end{equation}

Now let us complete our task to obtain a differential equation for $ \theta $ from the above expression.
Since $ \psi_s $ as given by \eqref{eq:psi} is a formal integration of the Helmholtz equation \eqref{eq:H}, it reduces in turn to the source term $ 4 \pi\,G\,\rho $ by an operation of the Helmholtz operator $ \triangle - m_s^2 $\,.
Note that the Helmholtz operators are commutable.
Thus, operating $ (\triangle_\xi - \mu_2^2)\,(\triangle_\xi - \mu_0^2) $ and using the Helmholtz equations for $ \psi_2 $ and $ \psi_0 $\,, we obtain a sixth-order differential equation for $ \theta $\,:
\begin{equation}
(\triangle_\xi - \mu_0^2)\,(\triangle_\xi - \mu_2^2)\,\triangle_\xi \theta
 + (\alpha_0\,\mu_0^2 + \alpha_2\,\mu_2^2)\,\triangle_\xi \theta^n
+ \mu_0^2\,\mu_2^2\,\theta^n
= 0\,.
\label{eq:master}
\end{equation}
This is our master differential equation for the non-relativistic stellar structure in HCG.
The fact that this is a sixth-order differential equation, as opposed to the second-order Lane--Emden equation in GR, is a direct consequence of the existence of three physical degrees of freedom in generic HCG.
For a derivation of \eqref{eq:master} via a different route, see \cite{Chen:2001a}.

\subsection{Limiting cases}

While we assumed finiteness of the graviton masses $ m_s^2 $'s in the derivation of the full equation \eqref{eq:master}, one would expect it contains various limiting cases corresponding to subclasses of HCG:
when either of the mass parameters $ \mu_2^2 $ or $ \mu_0^2 $ is taken to infinity, i.e., either of $ \alpha $ or $ \beta $ appearing in the HCG action \eqref{eq:action} vanishes, the gravity theory should reduce to ``$ R+R^2 $'' ($ \mu_2^2 \to \infty $) or ``$ R+C^2 $'' ($ \mu_0^2 \to \infty $), and when both of $ \mu_s^2 $ blow up, i.e., both of $ \alpha $ and $ \beta $ go to zero, the theory should recover GR.
In particular, in the last case, the standard LE equation (see below) should be reproduced somehow.
As we confirm shortly, this is indeed the case:
taking either or both of $ \mu_s^2$'s to infinity leads to correct variants of the LE equation in the corresponding subclasses of gravitational theories, although one should be fairly cautious when taking such limits since they lead to changes in the order of differentiation, reflecting the changes in the number of dynamical dofs.

As announced, when $ \mu_0^2 \to \infty $ and $ \mu_2^2 \to \infty $, the theory reduces to GR.
In this limit, only terms in proportion to the product $ \mu_0^2\,\mu_2^2 $ in \eqref{eq:master} should remain, reducing it to a second-order differential equation
\begin{equation}
\triangle_\xi \theta + \theta^n
= 0\,.
\label{eq:LE}
\end{equation}
This is nothing but the standard LE equation in the newtonian limit of GR.

Next, when only $ \mu_2^2 \to \infty $, the theory reduces to ``$ R+R^2 $'' gravity.
Then, retaining the terms in proportion to $ \mu_2^2 $ in \eqref{eq:master} leads to a fourth-order equation
\begin{equation}
(\triangle_\xi - \mu_0^2)\,\triangle_\xi \theta
- (\alpha_2\,\triangle_\xi + \mu_0^2)\,\theta^n
= 0\,.
\label{eq:master40}
\end{equation}
By recalling $ \alpha_2 = -4/3 $, one can straightforwardly show that the same equation derives by starting from the ``$ R+R^2 $'' action, i.e., \eqref{eq:action} with $ \alpha = 0 $.
This equation was first derived by Chen \textit{et al.} \cite{10.1143/PTP.106.63} in 2001, but their paper seems to have drawn little attention for long.
For example, the authors of Ref.~\cite{Capozziello:2011nr,Farinelli:2013pza} have made attempts to solve an equivalent but relatively more intricate integro-differential equation.

The other limit $ \mu_0^2 \to \infty $ corresponding to the ``$ R+C^2 $'' gravity reduces \eqref{eq:master} to a similar fourth-order equation
\begin{equation}
(\triangle_\xi - \mu_2^2)\,\triangle_\xi \theta
- (\alpha_0\,\triangle_\xi + \mu_2^2)\,\theta^n
= 0\,.
\end{equation}
To our best knowledge, this equation has not been obtained in the literature.
A crucial qualitative difference from \eqref{eq:master40} is the positive sign of the coefficient $ \alpha_0 = 1/3 $, which, as remarked before, determines whether the extra gravitational force is attractive or repulsive.

For later convenience, we shall write the above fourth-order equations in the common form
\begin{equation}
(\triangle_\xi - \mu_s^2)\,\triangle_\xi \theta
+ ((1 + \alpha_s)\,\triangle_\xi - \mu_s^2)\,\theta^n
= 0\,,
\label{eq:master4}
\end{equation}
where we have used $ \alpha_2 + \alpha_0 = -1 $.

\subsection{Stellar mass formulae}

It might be useful to give a formula for the total stellar mass $ M $ in terms of derivatives of the profile function $ \theta $ evaluated at the stellar surface.
By virtue of \eqref{eq:master}, $ M $ can be written as
\begin{equation}
\begin{aligned}
M
&
= 4\pi\,\ell^3\,\rho_\mathrm c\,\int_0^{\xi_R}\!\mathrm d\xi\,\xi^2\,\theta(\xi)^n \\
&
= -\frac{4\pi\,\ell^3\,\rho_\mathrm c}{\mu_2^2\,\mu_0^2}\,\xi_R^2\,
  \frac{\mathrm d}{\mathrm d\xi} \left[
   (\alpha_2\,\mu_2^2 + \alpha_0\,\mu_0^2)\,\theta^n
   + (\triangle_\xi-\mu_2^2)\,(\triangle_\xi-\mu_0^2)\,\theta
  \right]_{\xi_R}\,.
\end{aligned}
\end{equation}
In the fourth order limit, i.e., either $ \mu_0 $ or $ \mu_2 $ goes to $ \infty $\,, the expression reduces to
\begin{equation}
M
= \frac{4\pi\,\ell^3\,\rho_\mathrm c}{\mu_s^2}\,\xi_R^2\,
  \frac{\mathrm d}{\mathrm d\xi}\left[
   (\triangle_\xi-\mu_s^2)\,\theta + (1+\alpha_s)\theta^n
  \right]_{\xi_R}\,.
\end{equation}
In the GR limit, it recovers the standard formula
\begin{equation}
M
= - 4\pi\,\ell^3\,\rho_\mathrm c\,\xi_R^2\,
   \left.
   \frac{\mathrm d\theta}{\mathrm d\xi}
   \right|_{\xi_R}\,.
\label{eq:GRM(R)}
\end{equation}

\section{\label{sec:bc}Boundary conditions}

Now we move on to discuss boundary conditions for the profile function $ \theta $.
Since the master equation \eqref{eq:master} is sixth order in differentiation, there is need for six independent conditions, a priori.
We shall demonstrate all such conditions can be derived as the requirements for the compatibility with the hydrostatic equilibrium condition.
Here, we take the same strategy as in GR in the sense that we aim at imposing all the conditions on the values of $ \theta $ and its derivatives at the stellar center.
It will turn out, however, that this is made only partially successful by the massive nature of the extra gravitational potential.

Let us start by expressing the equilibrium condition \eqref{eq:hydro} in terms of $ \theta $ as
\begin{equation}
\theta'
= -\frac{\ell^{-1}}{4\pi\,G\,\rho_c}\,
  \frac{\mathrm d\Psi}{\mathrm dr}\,,
\label{eq:hydro2}
\end{equation}
where and hereafter the prime denotes differentiation with respect to $ \xi $.
In order for this condition to hold at a given point, one has to ensure all the derivatives of both sides match at that position.
Therefore, apart from the normalization $ \theta_\mathrm c = \theta(0) = 1 $, the derivatives at the stellar center $ \theta^{(n)}_\mathrm c \equiv \frac{\mathrm d^n\theta}{\mathrm d\xi^n}(0) $ are restricted to be consistent with the behavior of the potential $ \Psi = \phi + \alpha_2\,\psi_2 + \alpha_0\,\psi_0 $ there.
Indeed, one finds the expansion of the potentials to a sufficient order as
\begin{equation}
\begin{aligned}
G^{-1}\,\phi
&
= G^{-1}\,\phi(0)
  + \frac{2\pi}{3}\,\rho(0)\,r^2
  + \frac{\pi}{3}\,\frac{\mathrm d\rho}{\mathrm dr}(0)\,r^3
  + \frac{\pi}{10}\,\frac{\mathrm d^2\rho}{\mathrm dr^2}(0)\,r^4
  + \mathcal O(r^5)\,,\\
G^{-1}\,\psi_s
&
= -m_s\,I_s
  + \left(
     \frac{2 \pi}{3}\,\rho(0)
     - \frac{m_s^3\,I_s}{6}
    \right)\,
    r^2
  + \frac{\pi}{3}\,\frac{\mathrm d\rho}{\mathrm dr}(0)\,r^3  \\
& \quad
  + \left(
     \frac{\pi}{10}\,\frac{\mathrm d^2\rho}{\mathrm dr^2}(0)
     + \frac{\pi\,m_s^2\,\rho(0)}{30}
     - \frac{m_s^5\,I_s}{120}
    \right)\,r^4
  + \mathcal O(r^5)\,.
\end{aligned}
\end{equation}
From above, it is immediately seen that the radial acceleration at the stellar center vanishes, $ \lim_{r\to 0} -\frac{\mathrm d\Psi}{\mathrm dr} = 0 $, which implies $ \theta_\mathrm c' = 0 $ via \eqref{eq:hydro2}.
Hence, the following two boundary conditions have so far been obtained:
\begin{equation}
\theta_\mathrm c
= 1\,,
\quad
\theta'_\mathrm c
= 0\,.
\label{eq:bc_LE}
\end{equation}
These two conditions just suffice in the case of the second-order LE equation \eqref{eq:LE}.
In this case, all the higher derivatives at the stellar center are readily read off as
\begin{equation}
\theta''_\mathrm c
= -\frac{1}{3}\,,
\quad
\theta'''_\mathrm c
= 0\,,
\quad
\theta^{(4)}_\mathrm c
= \frac{n}{5}\,,
\quad
\theta^{(5)}_\mathrm c
= 0\,,
\quad
\cdots\,.
\end{equation}
By contrast, four more boundary conditions than \eqref{eq:bc_LE} are required in order to solve the full sixth-order differential equation \eqref{eq:master}.
They are found from derivatives of \eqref{eq:hydro2} as
\begin{equation}
\begin{aligned}
\theta''_\mathrm c
&
= -\frac{1}{4\pi\,G\,\rho_\mathrm c}\,
  \lim_{r \to 0} \frac{\mathrm d^2\Psi}{\mathrm dr^2}
= \frac{1}{3}\,\sum_{s=0,2} \alpha_s\,\mu_s^3\,\iota_s\,, \\
\theta'''_\mathrm c
&
= -\frac{\ell}{4\pi\,G\,\rho_\mathrm c}\,
  \lim_{r \to 0} \frac{\mathrm d^3\Psi}{\mathrm dr^3}
= 0\,, \\
\theta^{(4)}_\mathrm c
&
= -\frac{\ell^2}{4\pi\,G\,\rho_\mathrm c}\,
  \lim_{r \to 0} \frac{\mathrm d^4\Psi}{\mathrm dr^4}
= -\frac{1}{5}\,\sum_{s=0,2} \alpha_s\,\mu_s^2
   + \frac{1}{5}\,\sum_{s=0,2} \alpha_s\,\mu_s^5\,\iota_s\,, \\
\theta^{(5)}_\mathrm c
&
= -\frac{\ell^3}{4\pi\,G\,\rho_\mathrm c}\,
  \lim_{r \to 0} \frac{\mathrm d^5\Psi}{\mathrm dr^5}
= 0\,,
\end{aligned}
\label{eq:bc}
\end{equation}
where we have normalized the constant $ I_s $ appearing in the massive potential $ \psi_s $ to
\begin{equation}
\iota_s
\equiv
  \frac{I_s}{4\pi\,\ell^3\,\rho_\mathrm c}
= \frac{1}{\mu_s}\,\int_0^{\xi_R}\!\mathrm d\xi\,\xi\,\theta(\xi)^n\,\mathrm e^{-\mu_s \xi}
\label{eq:iota}
\end{equation}
and iteratively applied the conditions arising from lower derivatives to the higher ones.
It is observed that some quantities at the stellar center, $ \theta''_\mathrm c $ and $ \theta^{(4)}_\mathrm c $\,, are now related to the stellar global quantity $ I_s $\,, which was moreover introduced to guarantee flatness at spatial infinity.
Of course, the values of $ I_s $ are as yet undetermined since the profile function $ \theta $ has not been solved at this stage.
Therefore, these expressions of the boundary values should be considered merely formal;
The problem is not formulated as a simple initial value problem as in GR, but we will have to perform some matching procedure between the boundary values at the stellar center and the integrals of the solution over the whole domain.
Note that the stellar radius $ \xi_R $ is simultaneously determined by this procedure.

We also derive the boundary conditions for the fourth-order case \eqref{eq:master4}, where four boundary conditions are required in total, two of which are \eqref{eq:bc_LE}.
The remaining two are
\begin{equation}
\begin{aligned}
\theta''_\mathrm c
&
= -\frac{1}{4\pi\,G\,\rho_\mathrm c}\,
  \lim_{r \to 0} \frac{\mathrm d^2\Psi}{\mathrm dr^2}
= -\frac{1}{3}
  - \frac{\alpha_s}{3}\,(1 - \mu_s^3\,\iota_s)\,, \\
\theta'''_\mathrm c
&
= -\frac{\ell}{4\pi\,G\,\rho_\mathrm c}\,
  \lim_{r \to 0} \frac{\mathrm d^3\Psi}{\mathrm dr^3}
= 0\,.
\end{aligned}
\label{eq:bc4}
\end{equation}
Note that $ \theta''_\mathrm c $ differs from that for the sixth-order equation.

For comparison, we would like to clarify the differences between past studies and ours with respect to the boundary conditions.
In the full sixth-order case, Chen and Shao \cite{Chen:2001a} state that they imposed continuity of (an analogue of) the gravitational potential and its derivatives at the stellar surface $ r = R $.
In our construction, on the contrary, their (dis)continuity is already inherent in Eqs.~\eqref{eq:phi} and \eqref{eq:psi} as remarked at the end of Sec.~\ref{sec:bc}.
So, we have failed to find a direct analogue of our Eq.~\eqref{eq:bc} in their paper.
In the reduced fourth-order case, Chen \textit{et al.} \cite{10.1143/PTP.106.63} again mention continuity at the stellar surface, and we faced the same difficulty.
Capozziello \textit{et al.} \cite{Capozziello:2011nr} (and Farinelli \textit{et al.} \cite{Farinelli:2013pza}) are less talkative about the boundary conditions except for \eqref{eq:bc_LE}.

\section{\label{sec:sol}Solutions for $ n = 0 $ and $ 1 $}

In this section, we present analytical solutions for the polytropic indices $ n = 0 $ and $ 1 $ in each class of HCG.
While some of them are known in the literature, we present them in a more complete form clarifying every step of how the boundary conditions are used to determine integration constants.

Before moving on, we summarize the solutions to the LE equation \eqref{eq:LE} in GR, which should be recovered in the GR limit of HCG.
The exact solutions for $ n = 0,1 $ satisfying \eqref{eq:bc_LE} are well known:
\begin{equation}
\theta^\mathrm{LE}_0
= 1 - \frac{\xi^2}{6}\,,
\quad
\theta^\mathrm{LE}_1
= \frac{\sin \xi}{\xi}\,.
\label{eq:sol_LE}
\end{equation}
There is also an exact solution for $ n = 5 $, $ \theta^\mathrm{LE}_5 = 1/\sqrt{1+\frac{\xi^2}{3}} $, but this does not have a finite radius.
The stellar radius $ R $ in each case is
\begin{equation}
R_0
= \sqrt 6\,\ell_0\,,
\quad
R_1
= \pi\,\ell_1\,,
\end{equation}
where we have stressed here the length scale $ \ell $ depends on $ n $.
The masses and the internal potentials are respectively given by
\begin{equation}
M_0
= \frac{4}{3}\,\pi\,R_0^3\,\rho_c
= 8 \sqrt 6\,\pi\,\ell_0^3\,\rho_c\,,
\quad
M_1
= \frac{4}{\pi}\,R_1^3\,\rho_c
= 4 \pi^2\,\ell_1^3\,\rho_c\,,
\end{equation}
and
\begin{equation}
\phi_0(r<R_0)
= -\frac{G M_0}{R_0}\,\frac{3 R_0^2-r^2}{2 R_0^2}\,,
\quad
\phi_1(r<R_1)
= -\frac{G M_1}{R_1}\,\left(1 + \frac{\sin (\pi r/R_1)}{\pi r/R_1}\right)\,.
\end{equation}

\subsection{\label{sec:fourth}Fourth-order limit for ``$ R+R^2 $'' and ``$ R+C^2 $'' theories}

Next we discuss the cases where either $ \alpha $ or $ \beta $ is zero, for which the gravitational theory reduces to ``$ R + R^2 $'' ($ \alpha = 0 $) or ``$ R + C^2 $'' ($ \beta = 0 $), respectively.
Then the sixth-order master equation \eqref{eq:master} reduces to the fourth-order equation \eqref{eq:master4}.

\subsubsection{\label{sec:sol4n=0}$ n = 0 $}

For $ n = 0 $, Eq.~\eqref{eq:master4} reduces to an inhomogeneous linear equation
\begin{equation}
(\triangle_\xi - \mu_s^2)\,\triangle_\xi \theta
- \mu_s^2
= 0,
\label{eq:master4n=0}
\end{equation}
which depends on the mass parameter $ \mu_s $ but not on the coefficient $ \alpha_s $\,.
The solution, however, depends on the spin of the massive graviton, i.e., gravitational theory, as $ \alpha_s $ appears in the boundary conditions \eqref{eq:bc4}.
A procedure to find a general solution to equations of the type of \eqref{eq:master4n=0} is summarized in Appendix~\ref{sec:Helm}.
Following it, we find the general solution
\begin{equation}
\theta
= 1
  - \frac{\xi^2}{6}
  + A\,\frac{\sinh \mu_s \xi}{\xi}
  + B\,\frac{\cosh \mu_s \xi}{\xi}
  + C
  + \frac{D}{\xi}\,,
\end{equation}
where $ A, B, C, D $ are arbitrary constants of integration.
It may be interesting to note that Eq.~\eqref{eq:master4n=0} admits the LE solution in GR as a particular solution, but, as we will see shortly, it cannot satisfy the boundary conditions at the stellar center.

The ``LE'' boundary conditions \eqref{eq:bc_LE}, $ \theta_\mathrm c = 1 $ and $ \theta_\mathrm c' = 0 $, are so restrictive than they appear that three constants can be determined as $ B = D = 0 $ and $ C = -\mu_s\,A $\,, and we are left with the form with only one constant:
\begin{equation}
\theta
= 1
  - \frac{\xi^2}{6}
  + A\,\frac{\sinh \mu_s \xi - \mu_s\,\xi}{\xi}\,.
\end{equation}
One of the two extra conditions \eqref{eq:bc4}, $ \theta'''_\mathrm c = 0 $, has been already satisfied at this stage, and the last condition on $ \theta''_\mathrm c $ plays the role in fixing $ A $\,.
Indeed, the second derivative evaluated at the center $ \xi = 0 $ is
\begin{equation}
\theta''_\mathrm c
= -\frac{1}{3} + \frac{A\,\mu_s^3}{3}\,.
\end{equation}
Comparing this with \eqref{eq:bc4}, we find the relation between $ A $ and $ \iota_s $ as
\begin{equation}
A
= -\alpha_s\,(\mu_s^{-3} - \iota_s)\,.
\end{equation}
As we discussed in Sec.~\ref{sec:bc}, the value of $ \iota_s $ \eqref{eq:iota} in general involves an integral of the profile function over the stellar radius, which can never be determined before the solution is known.
In this sense, determination of $ A $ is subject to an appropriate matching procedure for the overall consistency.
In the case of the polytropic index $ n = 0 $, however, constancy of the stellar density, $ \rho = \rho_\mathrm c $\,, makes the matching procedure considerably simpler, as the integral $ \iota_s $ does not explicitly depend on $ \theta $.
Nonetheless, even in this case, evaluation of $ \iota_s $ is not trivial since the undetermined stellar radius $ \xi_R $ appears in its expression as
\begin{equation}
\iota_s
= \frac{1 - (\mu_s\xi_R+1)\,\mathrm e^{-\mu_s \xi_R}}{\mu_s^3}\,.
\end{equation}
At any rate, after eliminating $ A $\,, the solution satisfying all the boundary conditions is obtained as
\begin{equation}
\theta
= 1
  - \frac{\xi^2}{6}
  - \alpha_s\,\frac{(\mu_s\,\xi_R + 1)\,\mathrm e^{-\mu_s \xi_R}}{\mu_s^2}\,
    \frac{\sinh \mu_s \xi - \mu_s\,\xi}{\mu_s\,\xi}\,.
\label{eq:sol4n=0}
\end{equation}
This expression indicates there is always non-zero deviation from the $ n = 0 $ LE solution.

The above solution is still considered ``formal'' since the stellar radius $ \xi_R $ must be fixed by solving the consistency condition $ \theta(\xi_R) = 0 $, which cannot be done analytically even in this simplest case.
In each case of the limiting theories, ``$ R+C^2 $'' or ``$ R+R^2 $'', given the corresponding value of $ \alpha_s $\,, the radius $ \xi_R $ becomes a function of the mass parameter $ \mu_s $\,.
The solution in ``$ R+R^2 $'' gravity, with $ s = 0 $ and $ \alpha_0 = 1/3 $, is identical with Eq.~(30) of Ref.~\cite{Capozziello:2011nr},\footnote{There is a discrepancy between the numerical factor in Eq.~(28) with (31) of Ref.~\cite{10.1143/PTP.106.63} and ours.} while the solution in ``$ R+C^2 $'' gravity, with $ s = 2 $ and $ \alpha_2 = -4/3 $, seems to have been undiscovered in the literature.
Once $ \xi_R $ is determined, the stellar mass $ M $ and charge $ \Sigma_s $ can be evaluated via their expressions for $ n = 0 $:
\begin{equation}
M
= \frac{4\pi\,\ell^3\,\rho_\mathrm c\,\xi_R^3}{3}\,,
\quad
\Sigma_s
= 4 \pi\,\ell^3\,\rho_\mathrm c\,
  \frac{\mu_s\,\xi_R\,\cosh \mu_s \xi_R - \sinh \mu_s \xi_R}{\mu_s^3}\,.
\end{equation}
For the sake of completeness, we also present the gravitational potential inside the star
\begin{equation}
\Psi(r \leq R)
= -\frac{G M\,(3 R^2 - r^2)}{2R^3}
  - \alpha_s\,
    \frac{G \Sigma_s}{r}\,
    \frac{m_s\,r - (1+m_s\,R)\,\mathrm e^{-m_s R}\,\sinh m_s r}
         {m_s\,R\,\cosh m_s R - \sinh m_s R}\,.
\end{equation}

Before showing numerical results, let us overview some analytical properties of the solution \eqref{eq:sol4n=0}.
Considering $ \xi < \xi_R $\,, it reduces as $ \theta \to \theta_0^\mathrm{LE} = 1 - \xi^2/6 $ when the GR limit $ \mu_s^2 \to \infty $ is taken.
When the massless limit $ \mu_s^2 \to 0 $, at the opposite extreme, is taken, the profile function reduces as $ \theta \to 1 - (1+\alpha_s)\,\xi^2/6 $, where the coefficient $ \alpha_s $ plays a crucial role.
In ``$ R+R^2 $'' gravity, $ \alpha_0 = 1/3 $, the modification to the profile merely amounts to a moderate shrinkage of the stellar radius $ \xi_R $ from the Lane--Emden value of $ \sqrt 6 $ to $ 3/\sqrt 2 $, a decrease by a factor of $ 1/\sqrt{1+\alpha_0} = \sqrt 3/2 $.
This reflects the attractive nature of the spin-$ 0 $ graviton in HCG.
On the other hand, in ``$ R+C^2 $'' gravity, $ \alpha_2 = -4/3 $, the profile function no longer acquires a positive zero, failing to express a star with a finite radius.
This is explained by the repulsive nature of the spin-$ 2 $ graviton, whose strength exceeds that of the attractive force mediated by the ordinary massless graviton.
In reality, however, as long as these gravitons have a finite mass, their effects are restricted within a finite range $ r \lesssim m_s^{-1} $, and the stellar radius remains finite in any case.

\begin{figure}[htbp]
\begin{center}
\includegraphics[scale=.8]{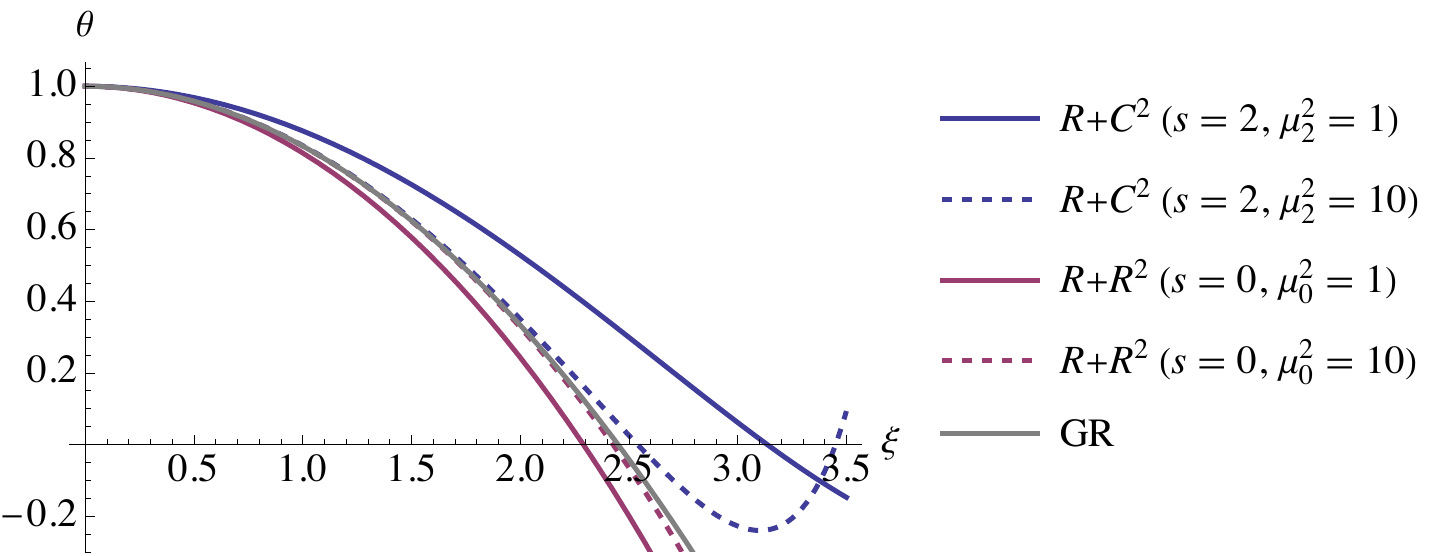}
\end{center}
\caption{\label{fig:sol4n=0}The profile function $ \theta $ in ``$ R + C^2 $'' gravity (blue) and ``$ R + R^2 $'' gravity (red) for the polytropic index $ n = 0 $. The massive spin-$ 2 $ (spin-$ 0 $) graviton provides a repulsive (attractive) force. As the graviton mass $ \mu_s^2 $ increases, the profile converges to the $ n = 0 $ LE solution in GR, $ \theta_0^\mathrm{LE} = 1 - \xi^2/6 $ (grey).}
\end{figure}

Figure~\ref{fig:sol4n=0} shows typical examples of the solution \eqref{eq:sol4n=0} in each gravity theory with different mass parameters $ \mu_s^2 $\,.
We see that the stellar radius in ``$ R+R^2 $'' gravity is smaller than the value in GR, whereas larger in ``$ R+C^2 $'' gravity.
This reveals that, as anticipated, the massive spin-$ 0 $ graviton arising from the addition of $ R^2 $ term provides an attractive force and the massive spin-$ 2 $ from the $ C^2 $ term provides a repulsive force.

\begin{figure}[htbp]
\includegraphics[scale=.8]{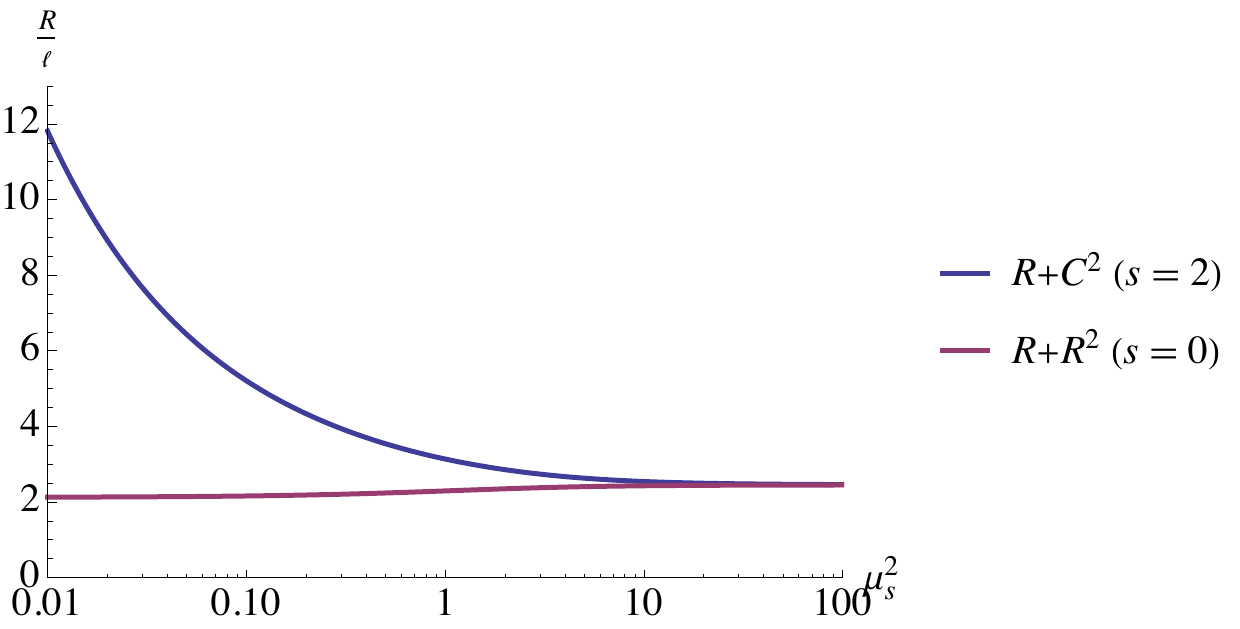}
\includegraphics[scale=.8]{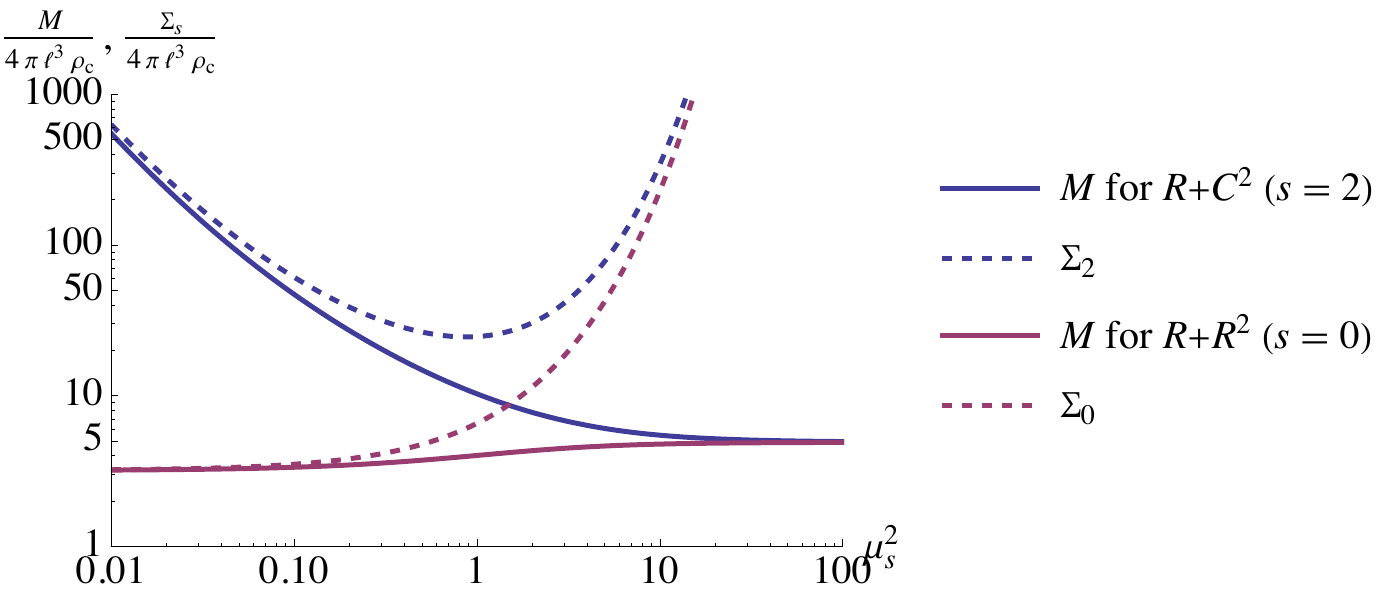}
\caption{\label{fig:RMS4n=0}The dependences of the stellar radius $ R $ (top), mass $ M $ (bottom, solid), and charge $ \Sigma_s $ (bottom, dashed), each appropriately normalized, on the graviton mass $ \mu_s^2 $ for the polytropic index $ n = 0 $. The stellar radius and mass are larger in ``$ R + C^2 $'' gravity (blue) and smaller in ``$ R + R^2 $'' gravity (red), but they both approach to the GR values of $ \sqrt 6 $ and $ 2 \sqrt 6 $, respectively, as the graviton mass $ \mu_s^2 $ increases. The charge $ \Sigma_s $ blows up in the GR limit, but the potential $ \psi_s $ then tends to $ 0 $. In the massless limit of ``$ R+R^2 $'' gravity, the limiting values are: $ R/\ell \to 3/\sqrt 2 $ and $ M \simeq \Sigma_s \to (\sqrt 3/2)^3\,M_0^\mathrm{LE} $. In ``$ R+C^2 $'' gravity, these quantities grow unboundedly as the spin-$ 2 $ graviton mass $ \mu_2^2 $ approaches to $ 0 $.}
\end{figure}

The modifications in the stellar global quantities, i.e., radius, mass, and charge, are depicted in Fig.~\ref{fig:RMS4n=0}.
The top panel shows the normalized stellar radius $ \xi_R = R/\ell $ against the mass parameter $ \mu_s^2 $\,.
When $ \mu_s^2 $ is increased, $ R/\ell $ quickly converges to the GR value $ R/\ell = \sqrt 6 $ in both gravity cases, as expected.
On the other hand, when $ \mu_s^2 $ is decreased, the difference between the natures of the two theories signifies:
in the massless limit of ``$ R+R^2 $'' gravity (red), the limiting value of the radius is finite, $ R/\ell \to 3/\sqrt 2 $, while in the same limit of ``$ R+C^2 $'' gravity (blue), in contrast, it blows up.
The bottom panel shows the dependences of the total mass $ M $ and the total charge $ \Sigma_s $, both appropriately normalized, on the graviton mass $ \mu_s^2 $\,.
When $ \mu_s^2 $ goes to infinity, $ M $ quickly converges to the values of GR, $ M/(4\pi\,\ell^3\,\rho_\mathrm c) = (R/\ell)^3/3 = 2 \sqrt 6 $ as expected, whereas $ \Sigma_s $ increases unboundedly.
This is phenomenologically not problematic because the observable gravitational potential $ \psi_s $ securely converges to $ 0 $ in this limit.
A similar behavior of the charge is observed in the study of neutron stars in \cite{Bonanno:2021zoy}.
In the massless limit of ``$ R + R^2 $'' gravity (red), the limiting values of the stellar mass and charge are $ M \simeq \Sigma_s \to (\sqrt 3/2)^3\,M_0^\mathrm{LE} $\,, while they both grow unboundedly in ``$ R + C^2 $'' gravity (blue) in accordance with the increase in the radius.

\begin{figure}[htbp]
\includegraphics[scale=.8]{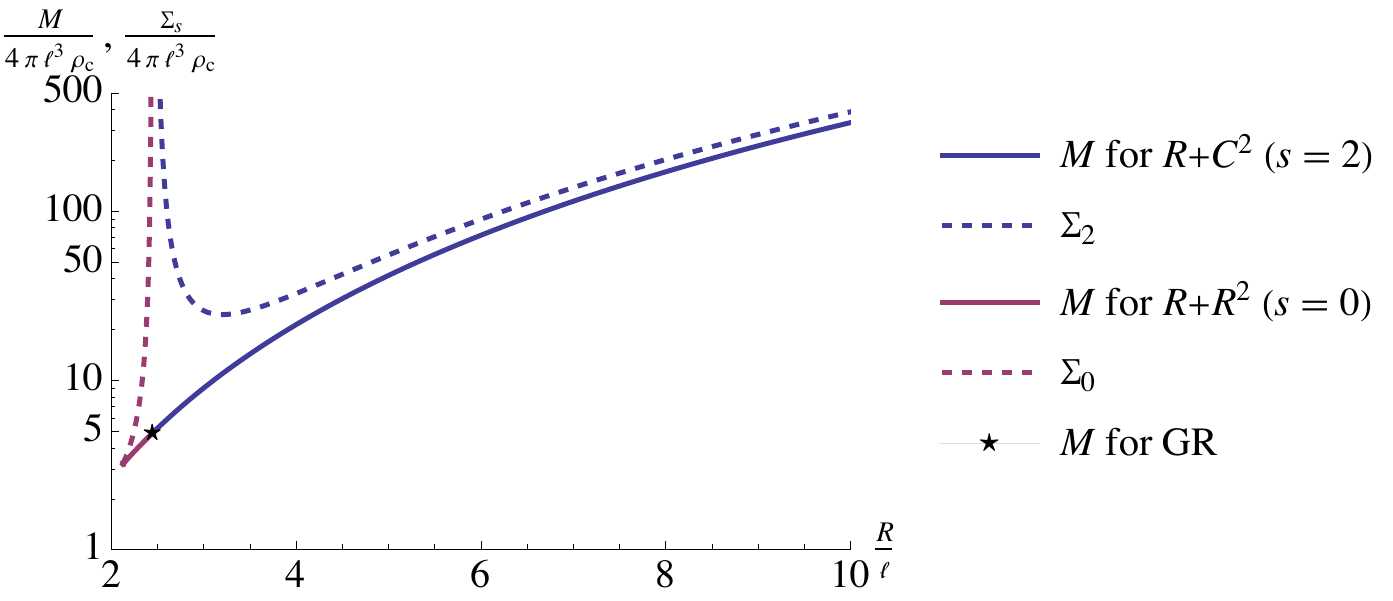}
\caption{\label{fig:MS-R4n=0}The normalized mass $ M/(4\pi\,\ell^3\,\rho_\mathrm c) $ (solid) and the normalized charge $ \Sigma_s/(4\pi\,\ell^3\,\rho_\mathrm c) $ (dashed) versus the normalized radius $ R/\ell $ for the polytropic index $ n = 0 $. The mass $ M $ and radius $ R $ in both gravity cases approach to the values in GR (black star) as the graviton mass $ \mu_s^2 $ increases.}
\end{figure}

Finally, Fig.~\ref{fig:MS-R4n=0} shows the relations between $ M $ and $ R $ (solid) and $ \Sigma_s $ and $ R $ (dashed), where the variables are appropriately normalized.
The $ M $--$ R $ curve reproduces the trivial relation for $ n = 0 $: $ M = \frac{4\pi}{3}\,R^3\,\rho_\mathrm c $\,.
$ \Sigma_s $ diverges when the radius approaches to the GR value $ R/\ell = \sqrt 6 $ as expected from the bottom panel of Fig.~\ref{fig:RMS4n=0}.
The fact that the value of $ \Sigma_s $ is comparable to $ M $ as long as $ \mu_s^2 \lesssim \mathcal O(1) $ implies that there is significant modification to the Newton law within distances shorter than $ m_s^{-1} $ from the stellar surface.

The radius $ R $ of a polytrope star is in proportion to the length scale $ \ell $ related to the physical conditions at the stellar center, such as the central density $ \rho_\mathrm c $ and pressure $ P_\mathrm c $\,, see \eqref{eq:ell}.
In this sense, there is a degeneracy between these physical conditions and gravity, including any possible modifications to GR, in measurements of stellar radius, for which it is incapable to quantify the effects of the massive gravitons independently of the properties and individual conditions of stellar matter.
Nevertheless, we here argue that, without going into direct comparisons with observational data, huge deviations in radius from the GR value with the same physical condition, as represented by $ \ell $, should be disfavored.
For instance, in ``$ R+C^2 $'' gravity, we may consider a radius $ R $ which is twice as large as the GR value, $ R_\mathrm{LE} = \sqrt 6\,\ell $, to be unlikely enough.
In order to have the ratio $ R/R_\mathrm{LE} \leq 2 $, the spin-$ 2 $ graviton mass must exceed $ 0.12 $, which can be interpreted as an upper bound on the parameter $ \sqrt\alpha/\ell = 1/\sqrt{2 \mu_2^2} < 2.0 $.

\subsubsection{\label{sec:sol4n=1}$ n = 1 $}

For the polytropic index $ n = 1 $, the fourth-order equation \eqref{eq:master4} reduces to a linear homogeneous equation
\begin{equation}
\triangle_\xi^2 \theta
+ (1 + \alpha_s - \mu_s^2)\,\triangle_\xi \theta
- \mu_s^2\,\theta
= 0\,.
\end{equation}
In order to find the solution, we ``factorize'' the differential operator to rewrite the above equation as
\begin{equation}
(\triangle_\xi + \lambda_+^2)\,(\triangle_\xi - \lambda_-^2)\,\theta
= 0
\end{equation}
with the ``roots''
\begin{equation}
\lambda_\pm
= \sqrt{\frac{\sqrt{(1 + \alpha_s - \mu_s^2)^2 + 4 \mu_s^2} \pm (1 + \alpha_s - \mu_s^2)}{2}}\,.
\end{equation}
It is obvious from the expression that $ \lambda_\pm $ are positive real irrespective of $ \alpha_s $ and $ \mu_s^2 $ (as long as $ \mu_s^2 > 0 $).
The general solution is a superposition of the fundamental solutions for the (homogeneous) Helmholtz equations with eigenvalues $ -\lambda_+^2 $ and $ \lambda_-^2 $, hence
\begin{equation}
\theta
= A_+\,\frac{\sin \lambda_+\,\xi}{\xi}
  + B_+\,\frac{\cos \lambda_+\,\xi}{\xi}
  + A_-\,\frac{\sinh \lambda_-\,\xi}{\xi}
  + B_-\,\frac{\cosh \lambda_-\,\xi}{\xi}\,.
\end{equation}

Let us determine the integration constants one by one.
By imposing the LE boundary conditions \eqref{eq:bc_LE}, $ \theta_\mathrm c = 1 $ and $ \theta_\mathrm c' = 0 $, three constants are fixed as $ B_+ = B_- = 0 $ and $ A_- = (1-A_+\lambda_+)/\lambda_- $\,.
Thus we find\footnote{A similar, but not identical expression is presented in Eq.~(38) of Ref.~\cite{10.1143/PTP.106.63}.}
\begin{equation}
\theta
= \frac{\sinh \lambda_-\,\xi
  + A_+\,(\lambda_-\,\sin \lambda_+\,\xi - \lambda_+\,\sinh \lambda_-\,\xi)}{\lambda_-\,\xi}\,.
\label{eq:theta4n=1}
\end{equation}
As in the case of $ n = 0 $, one more boundary condition $ \theta_\mathrm c''' = 0 $, being one of the remaining two in Eq.~\eqref{eq:bc4}, is already satisfied at this stage.
On the other hand, the yet unused second derivative $ \theta''_\mathrm c $ is given in terms of the constant $ A_+ $ as
\begin{equation}
\theta''_\mathrm c
= \frac{\lambda_-^2 - A_+\,\lambda_+\,(\lambda_+^2 + \lambda_-^2)}{3}\,.
\end{equation}
From \eqref{eq:bc4}, we find the relation between $ A_+ $ and the stellar global quantity $ \iota_s $ as
\begin{equation}
A_+
= \frac{\lambda_-^2 + 1 + \alpha_s\,(1 - \mu_s^3\,\iota_s)}
       {\lambda_+\,(\lambda_+^2+\lambda_-^2)}\,.
\label{eq:A+}
\end{equation}
Here, unlike the $ n = 0 $ case, $ \iota_s $ involves integration of $ \theta $, Eq.~\eqref{eq:theta4n=1}, so it necessarily contains the undetermined integration constant $ A_+ $\,.
Such an intermediate expression for $ \iota_s $ looks somewhat tedious, but has a simple linear (since $ n = 1 $) dependence on $ A_+ $\,:
\begin{equation}
\begin{aligned}
\iota_s
&
= \frac{\lambda_-
        - \mathrm e^{-\mu \xi_R}\,
          (\lambda_-\,\cosh \lambda_- \xi_R + \mu\,\sinh \lambda_- \xi_R)}
       {\lambda_-\,\mu_s\,(\mu_s^2 - \lambda_-^2)} \\
& \quad
  + \frac{A_+}
         {\lambda_-\,\mu_s\,(\mu_s^2 - \lambda_-^2)\,(\mu_s^2 + \lambda_+^2)}\,
    \biggl\{
          -\lambda_+\,\lambda_-\,(\lambda_+^2 + \lambda_-^2) \\
& \quad\quad
          + \mathrm e^{-\mu \xi_R}\,
            [\lambda_+\,(\mu_s^2 + \lambda_+^2)\,
             (\lambda_-\,\cosh \lambda_- \xi_R + \mu_s\,\sinh \lambda_- \xi_R)
             + \lambda_-\,(\lambda_-^2 - \mu_s^2)\,
               (\lambda_+\,\cos \lambda_+ \xi_R + \mu_s\,\sin \lambda_+ \xi_R)]
    \biggr\}\,.
\end{aligned}
\end{equation}
Substituting this into \eqref{eq:A+} gives back a linear equation for $ A_+ $\,, which can be explicitly solved as
\begin{equation}
A_+
= \left[
   \lambda_+
   + \frac{\lambda_-\,(\lambda_-^2 - \mu_s^2)\,
           (\lambda_+\,\cos \lambda_+ \xi_R + \mu_s\,\sin \lambda_+ \xi_R)}
          {(\mu_s^2 + \lambda_+^2)\,
           (\lambda_-\,\cosh \lambda_- \xi_R + \mu_s\,\sinh \lambda_- \xi_R)}
  \right]^{-1}\,,
\label{eq:const4n=1}
\end{equation}
where we have used the characteristic equations for $ \lambda_\pm $ to reduce the expression.
In this way, we have found the profile function $ \theta $ for $ n = 1 $ satisfying all the boundary conditions.
One should recall here that this expression is ``formal'' because it involves the stellar radius $ \xi_R $\,, which can only be found numerically by solving the consistency condition $ \theta(\xi_R) = 0 $.
Nonetheless, the expression ceases to contain $ \xi_R $ in the massive and massless limits.
In both gravity theories, the massive limit, $ \mu_s^2 \to \infty $, is the $ n = 1 $ LE solution in GR, $ \theta \to \theta_1^\mathrm{LE} = \xi^{-1}\,\sin \xi $.
On the other hand, in the massless limit, it reduces as $ \theta \to \sin(\sqrt{1+\alpha_s}\,\xi)/(\sqrt{1+\alpha_s}\,\xi) $.
This represents a rescaled LE solution for ``$ R+R^2 $'' gravity with $ \alpha_0 = 1/3 $, whereas it no longer has a finite radius for ``$ R+C^2 $'' gravity with $ \alpha_2 = -4/3 $.

\begin{figure}[htbp]
\includegraphics[scale=.8]{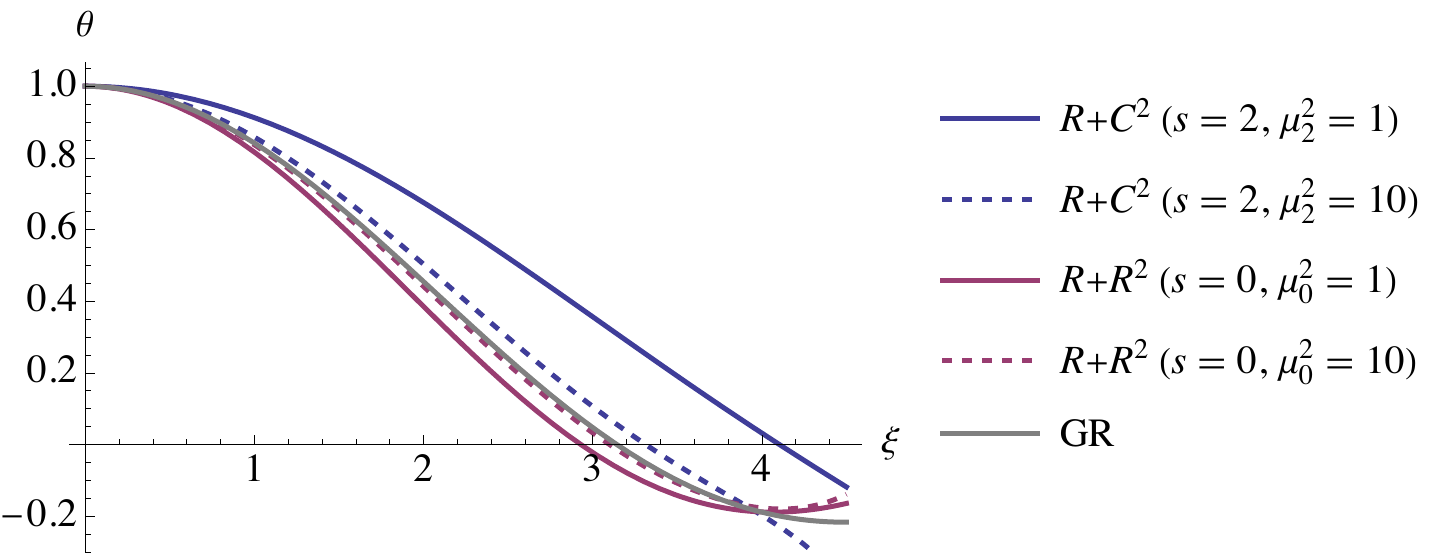}
\caption{\label{fig:sol4n=1}The solutions in ``$ R+C^2 $'' gravity (blue) and ``$ R+R^2 $'' gravity (red) together with the LE solution in GR (grey) for the polytropic index $ n = 1 $.}
\end{figure}

Figure \ref{fig:sol4n=1} compares the $ n = 1 $ solutions for ``$ R+C^2 $'' (blue) and ``$ R+R^2 $'' (red) theories with different values of $ \mu_s^2 $ with the $ n = 1 $ LE solution (grey).
As in the $ n = 0 $ case, in ``$ R+R^2 $'' (``$ R+C^2 $'') gravity, the radius becomes smaller (larger) than GR.
In the GR limit, $ \mu_s^2 \to \infty $, they all reduce to the LE solution.

\begin{figure}[htbp]
\includegraphics[scale=.8]{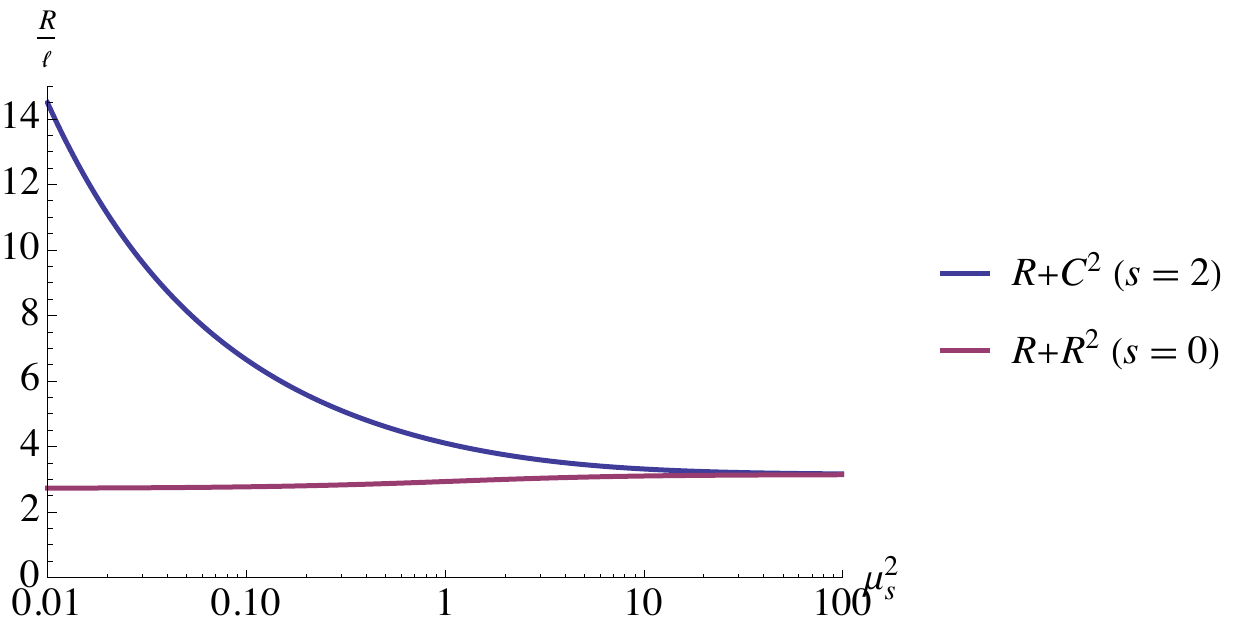}
\caption{\label{fig:R4n=1}The $ \mu_s $ dependences of the stellar radius $ R/\ell $ for the polytropic index $ n = 1 $.}
\end{figure}

The modification to the stellar radius is shown in Fig.~\ref{fig:R4n=1}, where we plot the values of the normalized radius $ \xi_R = R/\ell $ against the normalized mass parameter $ \mu_s $ for ``$ R+C^2 $'' (blue) and ``$ R+R^2 $'' (red).
Both curves converge to the GR value of $ \pi $ as $ \mu_s^2 $ blows up.
The massless limit for ``$ R+R^2 $'' gravity is $ R/\ell \to \sqrt 3\,\pi/2 $, whereas $ R/\ell $ increases unboundedly in ``$ R+C^2 $'' gravity as $ \mu_2^2 $ approaches to $ 0 $.

Unlike the $ n = 0 $ case, here again, the stellar mass $ M $ and charge $ \Sigma_s $ explicitly depend on $ \theta $, and hence one has to express them by substituting \eqref{eq:theta4n=1} together with \eqref{eq:const4n=1} into \eqref{eq:mass} and \eqref{eq:charge}, respectively, and evaluating them at the surface $ r = R $ using the numerical value of $ \xi_R $ for each choice of the mass parameter $ \mu_s^2 $\,.
Fortunately, in the current case, the integrations can be carried out analytically, giving explicit expressions for the mass and charge:
\begin{equation}
\begin{aligned}
M
&
= \frac{4\pi\,\ell^3\,\rho_\mathrm c}{\lambda_+^2 \lambda_-^3}\,
  \left[-A_+ \lambda_-^3 (\lambda_+\xi_R\cos{\lambda_+\xi_R}-\sin{\lambda_+\xi_R}) - \lambda_+^2(A_+\lambda_+ -1)(\lambda_-\xi_R\cosh{\lambda_-\xi_R}-\sinh{\lambda_-\xi_R})\right]\,,\\
\Sigma_s
&
= \frac{4\pi\,\ell^3\,\rho_\mathrm c\,\mathrm e^{\mu_s \xi_R}\,
        (\lambda_-\,\cosh \lambda_- \xi_R\,\sin \lambda_+ \xi_R
         - \lambda_+\,\sinh \lambda_- \xi_R\,\cos \lambda_+ \xi_R)}
       {\lambda_+\,(\lambda_+^2+\mu_s^2)\,
        (\lambda_-\,\cosh \lambda_- \xi_R + \mu_s\,\sinh \lambda_- \xi_R)
        + \lambda_-\,(\lambda_-^2-\mu_s^2)\,
          (\lambda_+\,\cos \lambda_+ \xi_R + \mu_s\,\sin \lambda_+ \xi_R)}\,,
\end{aligned}
\end{equation}
where we have taken advantage of maintaining $ A_+ $ in the expression of $ M $.
Similarly, one can also evaluate the gravitational potential $ \Psi = \phi + \alpha_s\,\psi_s $ inside a star by manipulating \eqref{eq:phi} and \eqref{eq:psi}, which also have analytical expressions in this case:
\begin{equation}
\begin{aligned}
\phi(r \leq R)
&
= -\frac{GM}{R}\,
  \left[
   1
   + \frac{R-r}{r}\,
     \frac{(A_+\,\lambda_+ - 1)\,\lambda_+^2\,
           (\lambda_-\,\xi\,\cosh \lambda_- \xi - \sinh \lambda_- \xi)
           + A_+\,\lambda_-^3\,(\lambda_+\,\xi\,\cos \lambda_+ \xi - \sin \lambda_+ \xi)}
          {(A_+\,\lambda_+ - 1)\,\lambda_+^2\,
           (\lambda_-\,\xi_R\,\cosh \lambda_- \xi_R - \sinh \lambda_- \xi_R)
           + A_+\,\lambda_-^3\,(\lambda_+\,\xi\,\cos \lambda_+ \xi_R - \sin \lambda_+ \xi_R)}
  \right]\,, \\
\psi_s(r \leq R)
&
= -\frac{\Sigma_s\,\mathrm e^{-m_s R}}{r}\,
  \frac{(\lambda_+\,\cos \lambda_+ \xi_R + \mu_s\,\sin \lambda_+ \xi_R)\,\sinh \lambda_- \xi
        - (\lambda_-\,\cosh \lambda_- \xi_R + \mu_s\,\sinh \lambda_- \xi_R)\,\sin \lambda_+ \xi}
       {\lambda_+\,\cos \lambda_+ \xi_R\,\sinh \lambda_- \xi_R
        - \lambda_-\,\sin \lambda_+ \xi_R\,\cosh \lambda_- \xi_R}\,.
\end{aligned}
\end{equation}

\begin{figure}[htbp]
\includegraphics[scale=.8]{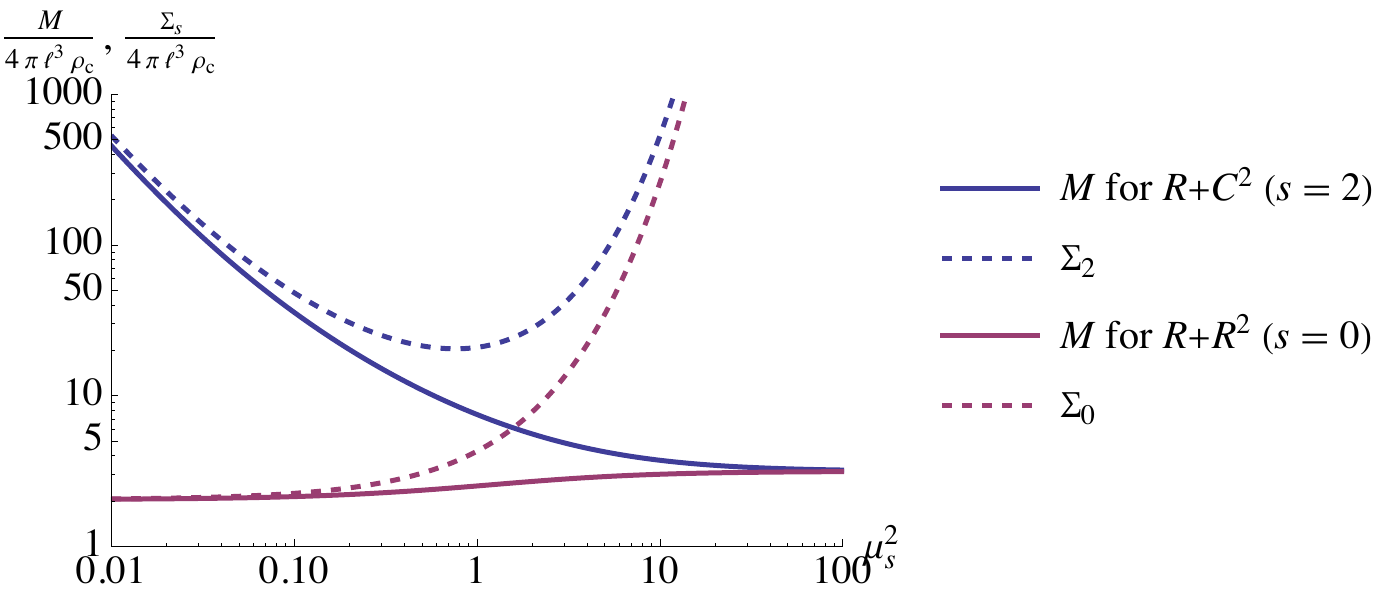}
\caption{\label{fig:MS4n=1}The $ \mu_s $ dependences of the stellar mass $ M $ (solid) and charge $ \Sigma_s $ (dashed) for the polytropic index $ n = 1 $.}
\end{figure}

Figure~\ref{fig:MS4n=1} shows the $ \mu_s $ dependences of $ M $ and $ \Sigma_s $\,.
The asymptotic value of $ M $ in the massive limit, $ \mu_s^2 \to \infty $, is the GR value of $ \pi $.
In the massless limit for ``$ R + R^2 $'' gravity (red), $ M $ and $ \Sigma_0 $ converge to the rescaled LE mass $ M \simeq \Sigma_0 \to (\sqrt 3/2)^3\,M_1^\mathrm{LE} $.
In contrast, both $ M $ and $ \Sigma_2 $ grow unboundedly for ``$ R + C^2 $'' gravity (blue) as $ \mu_2^2 \to 0 $.
These behaviors can be understood in a much similar fashion to the $ n = 0 $ case.

\begin{figure}[htbp]
\includegraphics[scale=.8]{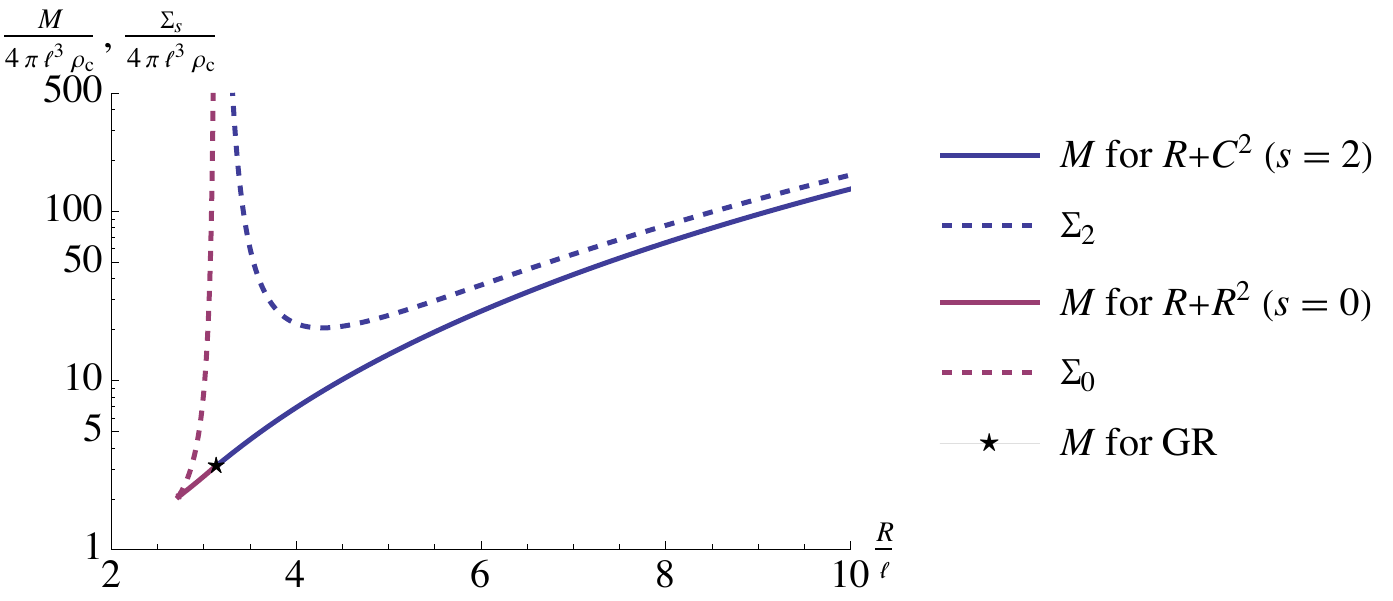}
\caption{\label{fig:MS-R4n=1}The $ M $--$ R $ (solid) and $ \Sigma_s $--$ R $ (dashed) relations for the polytropic index $ n = 1 $.}
\end{figure}

Finally, Fig.~\ref{fig:MS-R4n=1} shows the $ M $--$ R $ (solid) and $ \Sigma_s $--$ R $ (dashed) relations, where the quantities are appropriately normalized.
Similar trends to the $ n = 0 $ case show up in each diagram.

In the case of ``$ R+C^2 $'' gravity, from an argument that $ n = 1 $ polytrope stars should not acquire a radius and mass much larger than those in GR, we can place a reasonable lower bound on the spin-$ 2 $ graviton mass.
For instance, in order to have $ R/R_\mathrm{LE} < 2 $, we obtain $ \mu_2^2 > 0.12 $.
This can be converted into an upper bound on the theory parameter $ \alpha $ in the ``$ R+C^2 $'' action: $ \sqrt\alpha < 2.0\,\ell $.

\subsection{\label{sec:sixth}Full sixth-order equation in generic HCG}

Now we would like to tackle the full master equation \eqref{eq:master} in generic HCG.
The co-existence of the two extra dofs of spin-$ 2 $ and -$ 0 $ renders the analysis considerably messy, but most features of the solutions will be reasonably understood as collective contributions from the spin-$ 2 $ and -$ 0 $ dofs.
Indeed, in most occasions treated in this paper, either of the dofs dominates and the obtained solution therefore mimics some of those appeared in the previous fourth-order cases.

\subsubsection{$ n = 0 $}

In the case of $ n = 0 $, the master equation \eqref{eq:master} reduces to an inhomogeneous linear equation
\begin{equation}
(\triangle_\xi - \mu_0^2)\,(\triangle_\xi - \mu_2^2)\,\triangle_\xi \theta
+ \mu_0^2\,\mu_2^2
= 0\,.
\label{eq:mastern=0}
\end{equation}
The general solutions with six arbitrary constants are found following the procedure in Appendix~\ref{sec:Helm}.
For non-degenerate eigenvalues $ \mu_0 \neq \mu_2 $\,, it is
\begin{equation}
\theta
= 1
  - \frac{\xi^2}{6}
  + A_0\,\frac{\sinh \mu_0\,\xi}{\xi}
  + B_0\,\frac{\cosh \mu_0\,\xi}{\xi}
  + A_2\,\frac{\sinh \mu_2\,\xi}{\xi}
  + B_2\,\frac{\cosh \mu_2\,\xi}{\xi}
  + C
  + \frac{D}{\xi}\,.
\label{eq:gsoln=0}
\end{equation}
Although we are not so much concerned with the degenerate case $ \mu_0 = \mu_2 \equiv \mu $\,, the general solution in such a special case is
\begin{equation}
\theta
= 1
  - \frac{\xi^2}{6}
  + A\,\frac{\sinh \mu\,\xi}{\xi}
  + B\,\frac{\cosh \mu\,\xi}{\xi}
  + \tilde A\,\sinh \mu\,\xi
  + \tilde B\,\cosh \mu\,\xi
  + C
  + \frac{D}{\xi}\,.
\end{equation}
As in the fourth-order case, the $ n = 0 $ LE solution is again a particular solution but it will turn out not to satisfy the boundary conditions.

We shall concentrate on the non-degenerate case \eqref{eq:gsoln=0}.
This time we are to impose six boundary conditions in total as given by \eqref{eq:bc_LE} and \eqref{eq:bc}
By imposing first the LE boundary condition \eqref{eq:bc_LE}, we can fix three constants as $ C = -\mu_0\,A_0 - \mu_2\,A_2 $ and $ D = B_0 = B_2 = 0 $\,, and get a reduced form of the solution
\begin{equation}
\theta
= 1
  - \frac{\xi^2}{6}
  + A_0\,\frac{\sinh \mu_0\,\xi - \mu_0\,\xi}{\xi}
  + A_2\,\frac{\sinh \mu_2\,\xi - \mu_2\,\xi}{\xi}\,.
\end{equation}
At this point, the above solution already satisfies two of the four extra conditions in \eqref{eq:bc}, $ \theta'''_\mathrm c = \theta^{(5)}_\mathrm c = 0 $, and we are left with the requirements for $ \theta''_\mathrm c $ and $ \theta^{(4)}_\mathrm c $\,.
These derivatives are written in terms of the remaining constants $ A_0 $ and $ A_2 $ as
\begin{equation}
\theta_\mathrm c''
= -\frac{1 - A_0\,\mu_0^3 - A_2\,\mu_2^3}{3}\,,
\quad
\theta^{(4)}_\mathrm c
= \frac{A_0\,\mu_0^5 + A_2\,\mu_2^5}{5}\,.
\end{equation}
Then from \eqref{eq:bc}, $ A_0 $ and $ A_2 $ are related to the stellar integrals $ \iota_0 $ and $ \iota_2 $ as
\begin{equation}
A_0
= -\alpha_0\,(\mu_0^{-3} - \iota_0)\,,
\quad
A_2
= -\alpha_2\,(\mu_2^{-3} - \iota_2)\,,
\end{equation}
respectively.
Thanks to the constancy of $ \rho $ for the polytropic index $ n = 0 $, $ \iota_s $ are found the same, being independent of $ A_0 $ or $ A_2 $\,, as in the fourth-order case,
\begin{equation}
\iota_s
= \frac{1-(\mu_s\,\xi_R+1)\,\mathrm e^{-\mu_s \xi_R}}{\mu_s^3}\,,
\end{equation}
thereby fixing the remaining constants as
\begin{equation}
A_0
= -\alpha_0\,\frac{(\mu_0\,\xi_R+1)\,\mathrm e^{-\mu_0 \xi_R}}{\mu_0^3}\,,
\quad
A_2
= -\alpha_2\,\frac{(\mu_2\,\xi_R+1)\,\mathrm e^{-\mu_2 \xi_R}}{\mu_2^3}\,.
\end{equation}
Therefore, we get the solution satisfying all the boundary conditions:\footnote{We find this differs from Eq.~(30) with (33) of Ref.~\cite{Chen:2001a}.}
\begin{equation}
\theta
= 1
  - \frac{\xi^2}{6}
  + \alpha_0\,
    \frac{(\mu_0\,\xi_R+1)\,\mathrm e^{-\mu_0 \xi_R}}{\mu_0^3}\,
    \frac{\sinh \mu_0\,\xi - \mu_0\,\xi}{\xi}
  + \alpha_2\,
    \frac{(\mu_2\,\xi_R+1)\,\mathrm e^{-\mu_2 \xi_R}}{\mu_2^3}\,
    \frac{\sinh \mu_2\,\xi - \mu_2\,\xi}{\xi}\,.
\label{eq:thetan=0}
\end{equation}
The remaining parameter $ \xi_R $ is numerically determined by solving the consistency condition $ \theta(\xi_R) = 0 $ for given mass parameters $ \mu_2 $ and $ \mu_0 $\,.
After all, it is clear from the above expression that the gravitational effects from individual dofs are separated and purely additive in this case.
Moreover, due to the specialness of the $ n = 0 $ eos, where the mass density $ \rho $ is constant, the analytical expressions of the total stellar mass $ M $ and two charges $ \Sigma_2 $ and $ \Sigma_0 $ are identical with the ones in the fourth-order case:
\begin{equation}
M
= \frac{4\pi\,\ell^3\,\rho_\mathrm c\,\xi_R^3}{3}\,,
\quad
\Sigma_s
= 4 \pi\,\ell^3\,\rho_\mathrm c\,
  \frac{\mu_s\,\xi_R\,\cosh \mu_s \xi_R - \sinh \mu_s \xi_R}{\mu_s^3}\,.
\end{equation}
The gravitational potential inside a star is then found as
\begin{equation}
\Psi(r\leq R)
= -\frac{GM\,(3 R^2 - r^2)}{2R^3}
  - \sum_{s=0,2}
    \alpha_s\,
    \frac{G \Sigma_s}{r}\,
    \frac{m_s\,r - (1+m_s\,R)\,\mathrm e^{-m_s R}\,\sinh m_s r}
         {m_s\,R\,\cosh m_s R - \sinh m_s R}\,.
\end{equation}
Various massive and massless limits of the solution \eqref{eq:thetan=0} can be understood from the properties of the fourth-order $ n = 0 $ solutions discussed in Sec.~\ref{sec:sol4n=0}.
Among others, the spurious double massless limit $ \theta \to 1 $ seems to reflect some profound aspect of the full purely quadratic gravity.

\begin{figure}[htbp]
\includegraphics[scale=.8]{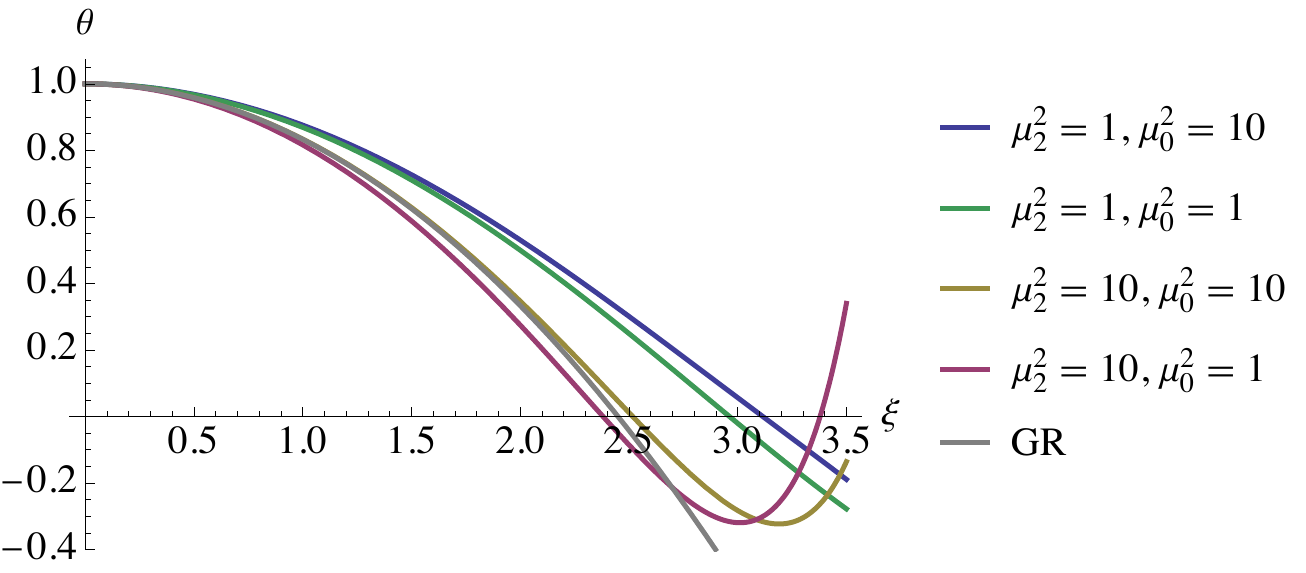}
\caption{\label{fig:soln=0}Examples of the profile functions for the polytropic index $ n = 0 $ compared with the {LE} solution (grey).}
\end{figure}

Some examples of the solution are shown in Fig.~\ref{fig:soln=0} together with the $ n = 0 $ {LE} solution $ \theta_0^\mathrm{LE} = 1-\xi^2/6 $ (grey).
The general tendency is that the lower the spin-$ 2 $ (spin-$ 0 $) graviton mass is, the more effectively the repulsive (attractive) force works.
Quantitatively, though, there is a significant difference between these two graviton effects, which shows up representatively in the case of $ \mu_2^2 = \mu_0^2 = 1 $ (green):
repulsion by spin-$ 2 $ graviton is much more noticeable than attraction by spin-$ 0 $, which was observed as well in the study of neutron stars \cite{Bonanno:2021zoy}.
This could be understood as a direct consequence of the ratio of the coefficients being $ \alpha_2/\alpha_0 = -4 $:
in the case of comparable graviton masses, $ \mu_2^2 \approx \mu_0^2 $\,, the influence coming from the massive spin-$ 2 $ graviton is four-fold stronger in magnitude compared to spin-$ 0 $.
Moreover, as the spin-$ 2 $ mass $ \mu_2^2 $ decreases below $ \mathcal O(1) $, the repulsive force can even overcome the attractive force of the massless graviton so that a star can puff up unboundedly, whereas the spin-$ 0 $ attractive force can merely strengthen gravity by at most a factor of a few, resulting in a bounded shrinkage of a star.

\begin{figure}[htbp]
\includegraphics[scale=.8]{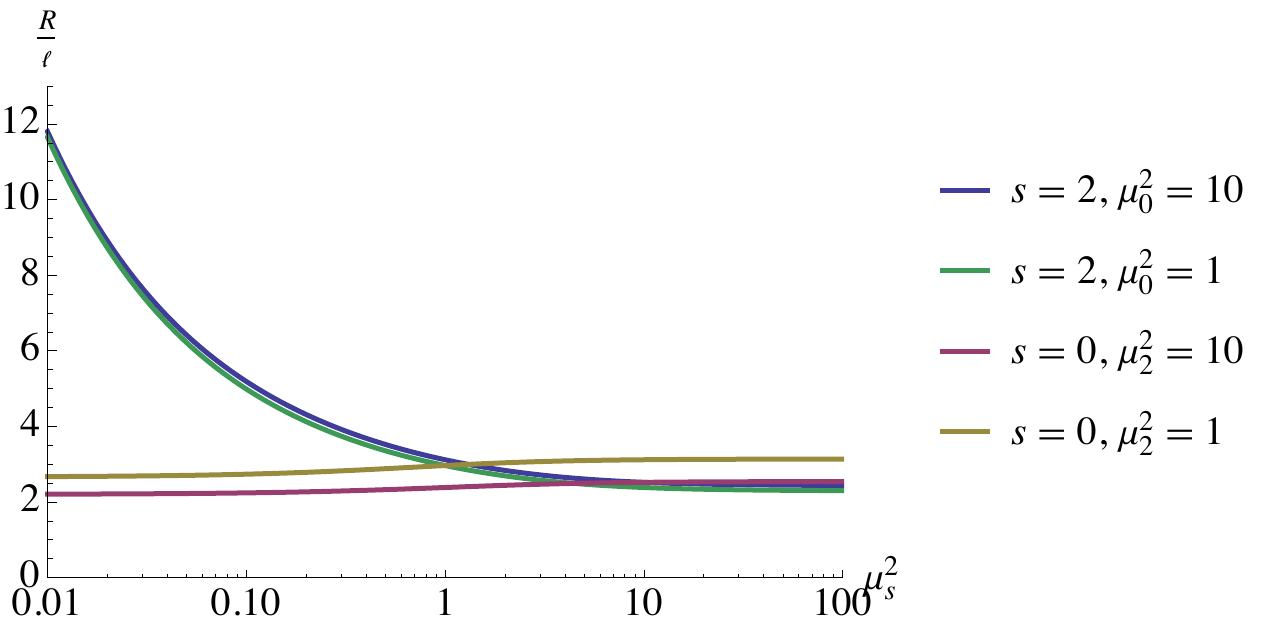}
\caption{\label{fig:Rn=0}The $ \mu_s^2 $ dependences of the normalized radius $ R/\ell $ for the polytropic index $ n = 0 $.}
\end{figure}

By solving the consistency condition $ \theta(\xi_R) = 0 $ numerically, we have the dependence of the normalized radius $ R/\ell $ on the mass parameters, some examples being plotted in Fig.~\ref{fig:Rn=0}.
In the figure, either of the two masses are varied while the rest is fixed.
The massive limit in this case corresponds to either of the reduced theories, ``$ R+R^2 $'' or ``$ R+C^2 $'', so the radius does not converge to the GR value of $ \sqrt 6 $;
It is only realized when the both masses are taken to infinity.
The radius remains finite in the massless limit of spin-$ 0 $ (red and yellow), whereas it blows up as the spin-$ 2 $ graviton mass approaches to $ 0 $ (blue and green).

\begin{figure}[htbp]
\includegraphics[scale=.8]{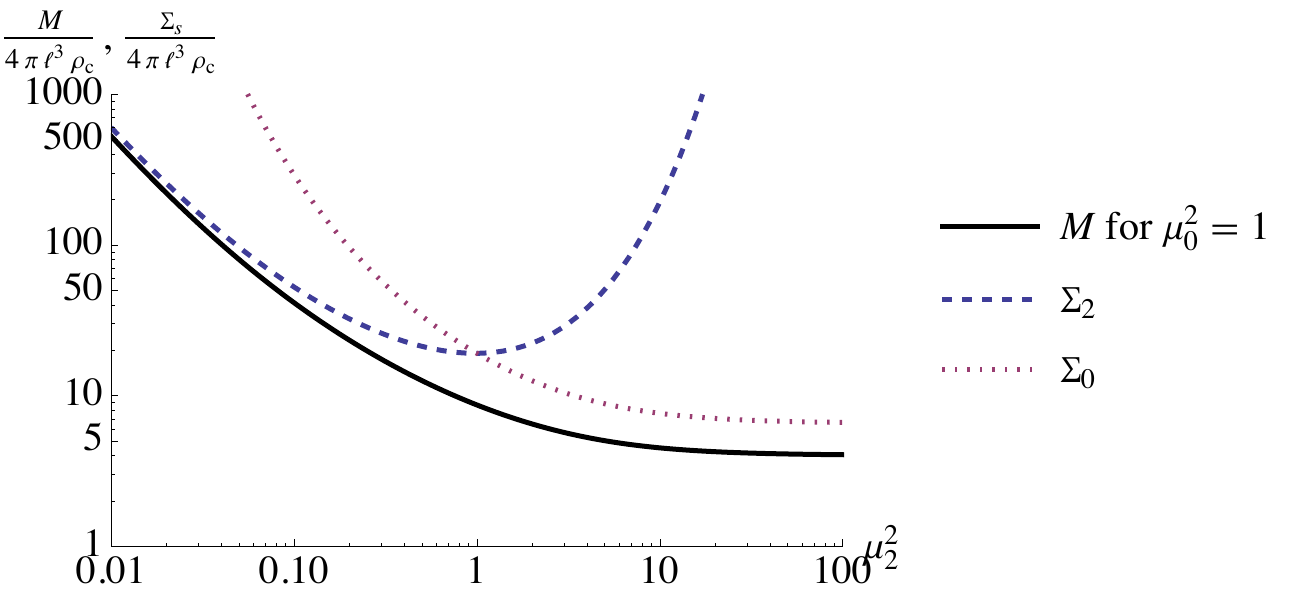}
\includegraphics[scale=.8]{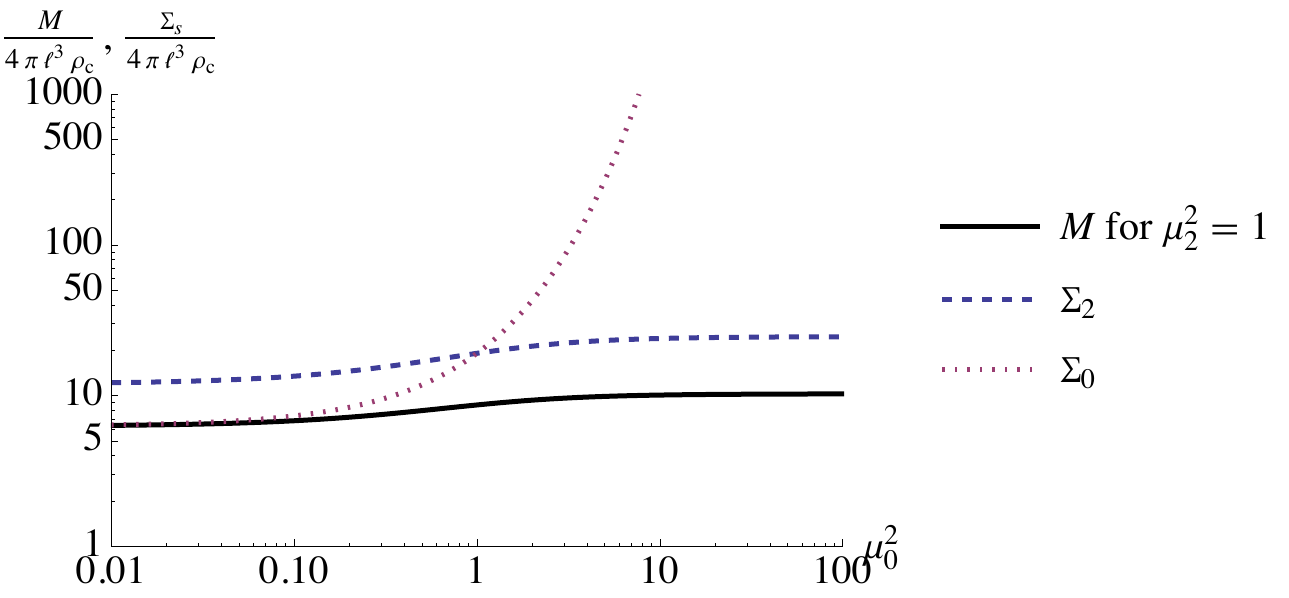}
\caption{\label{fig:MSSn=0}Typical $ \mu_s $ dependences of $ M $ (solid black), $ \Sigma_2 $ (dashed blue), and $ \Sigma_0 $ (dotted red) for $ n = 0 $. In the top (bottom) panel, $ \mu_2 $ ($ \mu_0 $) is varied while the other is fixed.}
\end{figure}

Figure~\ref{fig:MSSn=0} shows typical dependences on the graviton mass $ \mu_s $ of the stellar mass $ M $ and charges $ \Sigma_s $\,, where they are appropriately normalized.
In the top (bottom) panel, $ \mu_2^2 $ ($ \mu_0^2 $) is varied while $ \mu_0^2 $ ($ \mu_2^2 $) is fixed to $ 1 $.
The behavior of $ M $ can be understood in a similar way to the radius.
As for the charges, when the spin-$ s $ graviton mass $ \mu_s^2 $ goes to infinity, the corresponding spin-$ s $ charge $ \Sigma_s $ diverges, while the other charge $ \Sigma_{s'} $ ($ s' \neq s $) remains finite.
These divergences do not matter because the potential $ \psi_s $ vanishes in the massive limits.
On the other hand, in the massless limit of the spin-$ s $ graviton, the spin-$ s $ charge $ \Sigma_s $ tends to the mass $ M $.
The other charge $ \Sigma_{s'} $ has similar tendency as it is correlated with the stellar mass $ M $.

\begin{figure}[htbp]
\includegraphics[scale=.8]{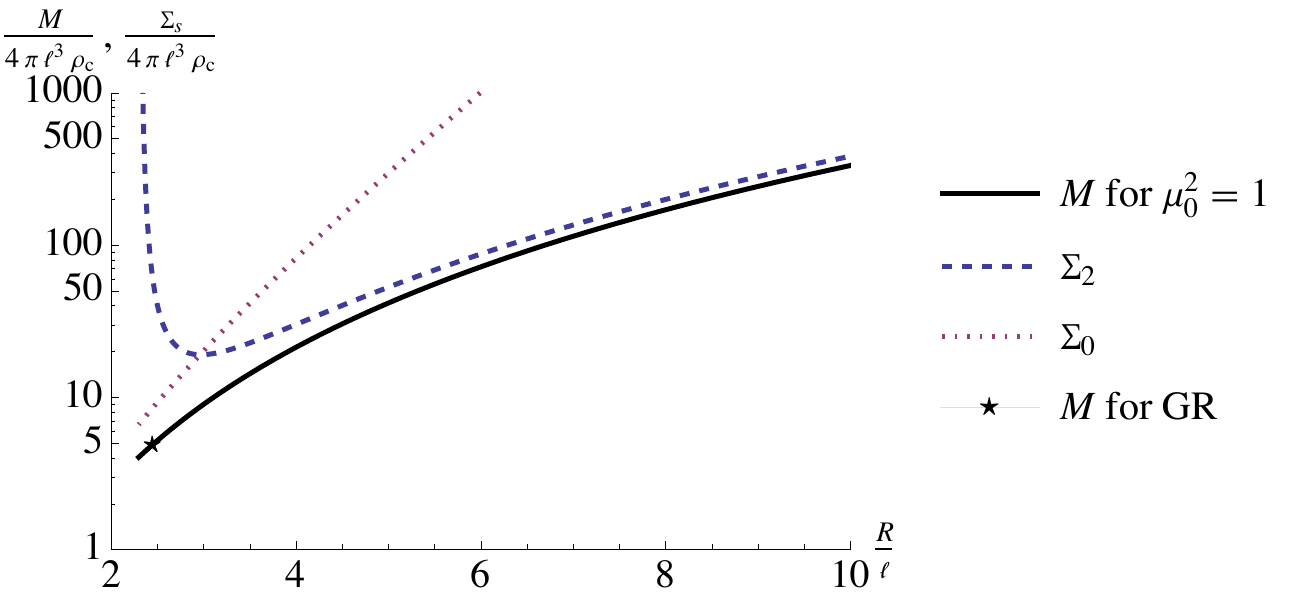}
\includegraphics[scale=.8]{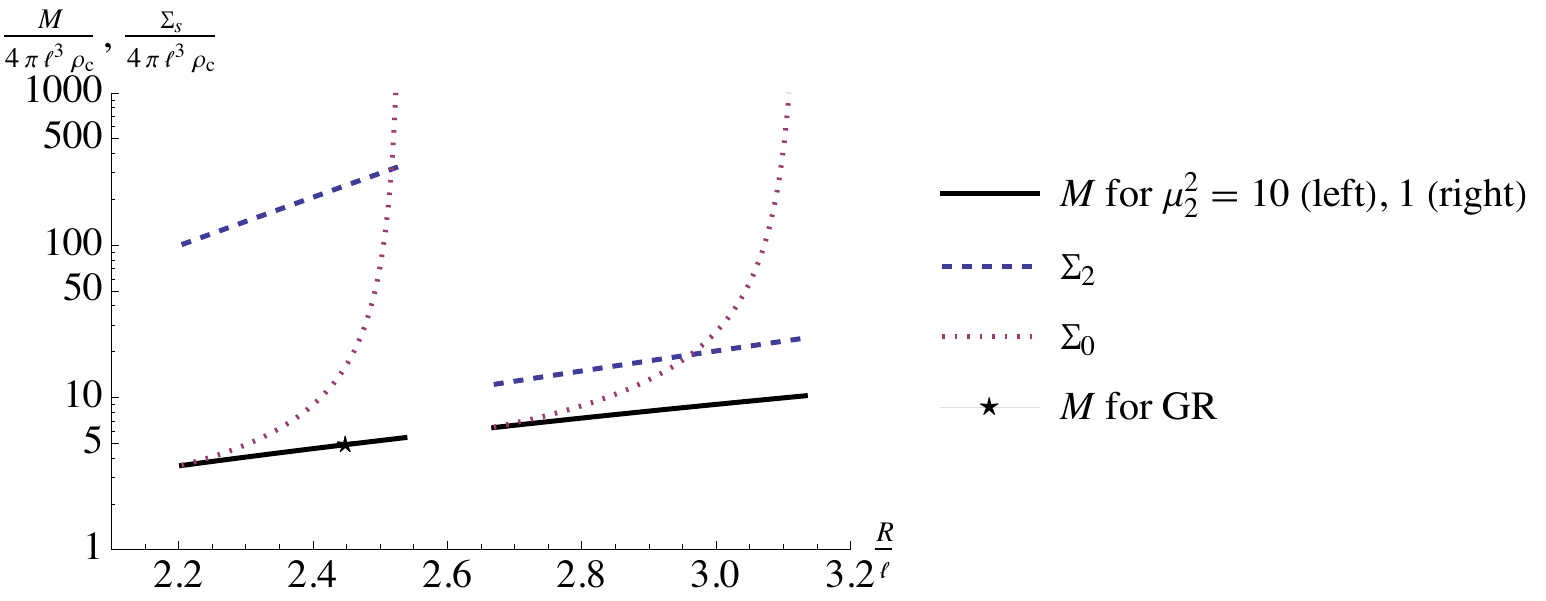}
\caption{\label{fig:MSS-Rn=0}Typical relations between $ M $ and $ R $ (solid black) and $ \Sigma_s $ and $ R $ (dashed blue for $ \Sigma_2 $ and dotted red for $ \Sigma_0 $) for $ n = 0 $. In the top panel, $ \mu_2^2 $ is varied while $ \mu_0^2 $ is fixed to $ 1 $. In the bottom panel, $ \mu_0^2 $ is varied while $ \mu_2^2 $ is fixed to $ 10 $ (left) or $ 1 $ (right).}
\end{figure}

After all, each panel in Fig.~\ref{fig:MSS-Rn=0} shows typical relations between $ M $ and $ R $ (solid black) and $ \Sigma_s $ and $ R $ (dashed blue and dotted red), where the values are appropriately normalized.
In the top (bottom) panel, $ \mu_2^2 $ ($ \mu_0^2 $) is varied while $ \mu_0^2 $ ($ \mu_2^2 $) is fixed.
One can confirm from the top panel that the possible ranges of the stellar radius $ R $ and mass $ M $ for varying $ \mu_2^2 $ are enormously large.
On the other hand, for a given value of $ \mu_2^2 $, the stellar radius and mass can only vary within a rather tiny range as seen in the bottom panel.

\begin{figure}[htbp]
\includegraphics[scale=.8]{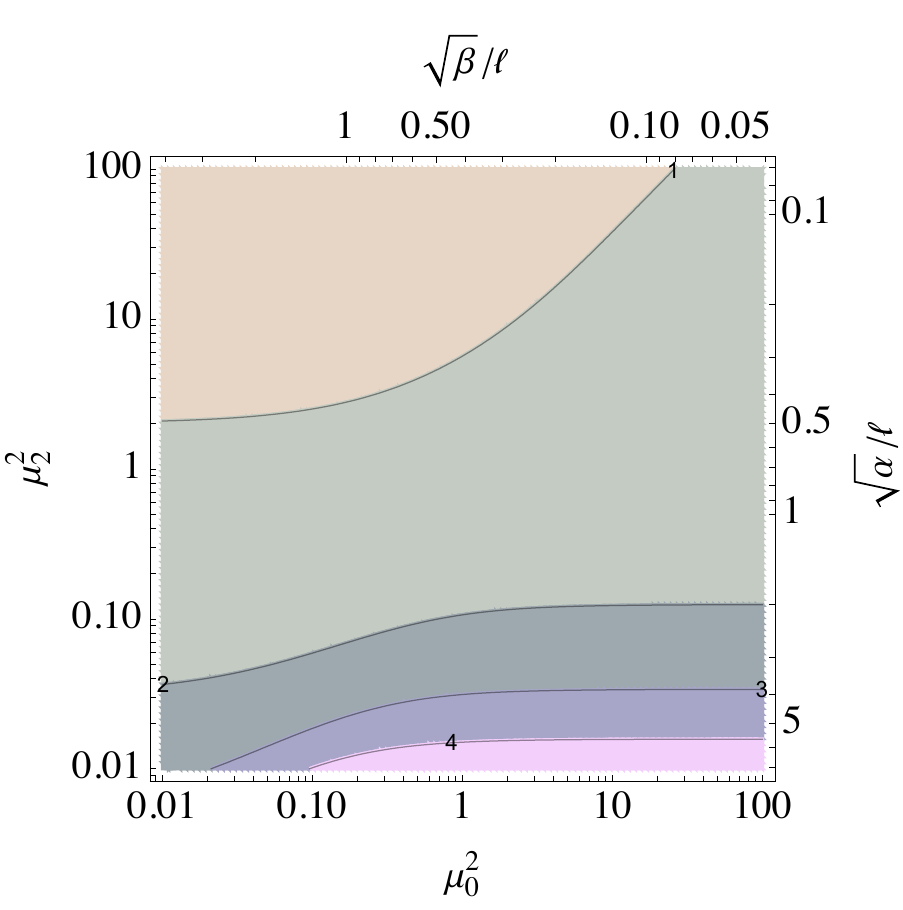}
\caption{\label{fig:ctn=0}Contours of the stellar radius $ R $ in the $ (\mu_0^2,\mu_2^2) $ plane  for the polytropic index $ n = 0 $. On the contours from top to bottom, the ratio of the calculated stellar radius to the GR value, $ R/R_\mathrm{LE} $, is $ 1,2,3,4 $.}
\end{figure}

Figure~\ref{fig:ctn=0} shows contours of the stellar radius $ R $ in the parameter plane $ (\mu_0^2,\mu_2^2) $, where on each contour, $ R $ has a multiple of the GR value $ R_\mathrm{LE} = \sqrt 6\,\ell $.
By demanding any $ n = 0 $ polytrope stars in the universe to have a radius no larger than some multiple, say $ 2 $, of the GR value, one finds a constraint on the combination of the theory parameters $ (\mu_0^2,\mu_2^2) $, or equivalently $ (\alpha,\beta) $.
Since the radius is sensitive to $ \mu_2^2 $ for $ R \gtrsim R_\mathrm{LE} $\,, $ \sqrt\alpha $ is generally constrained to below a few $ \ell $, whereas $ \sqrt\beta $ is virtually not restricted.

\subsubsection{$ n = 1 $}

\begin{figure}[htbp]
\includegraphics[scale=.8]{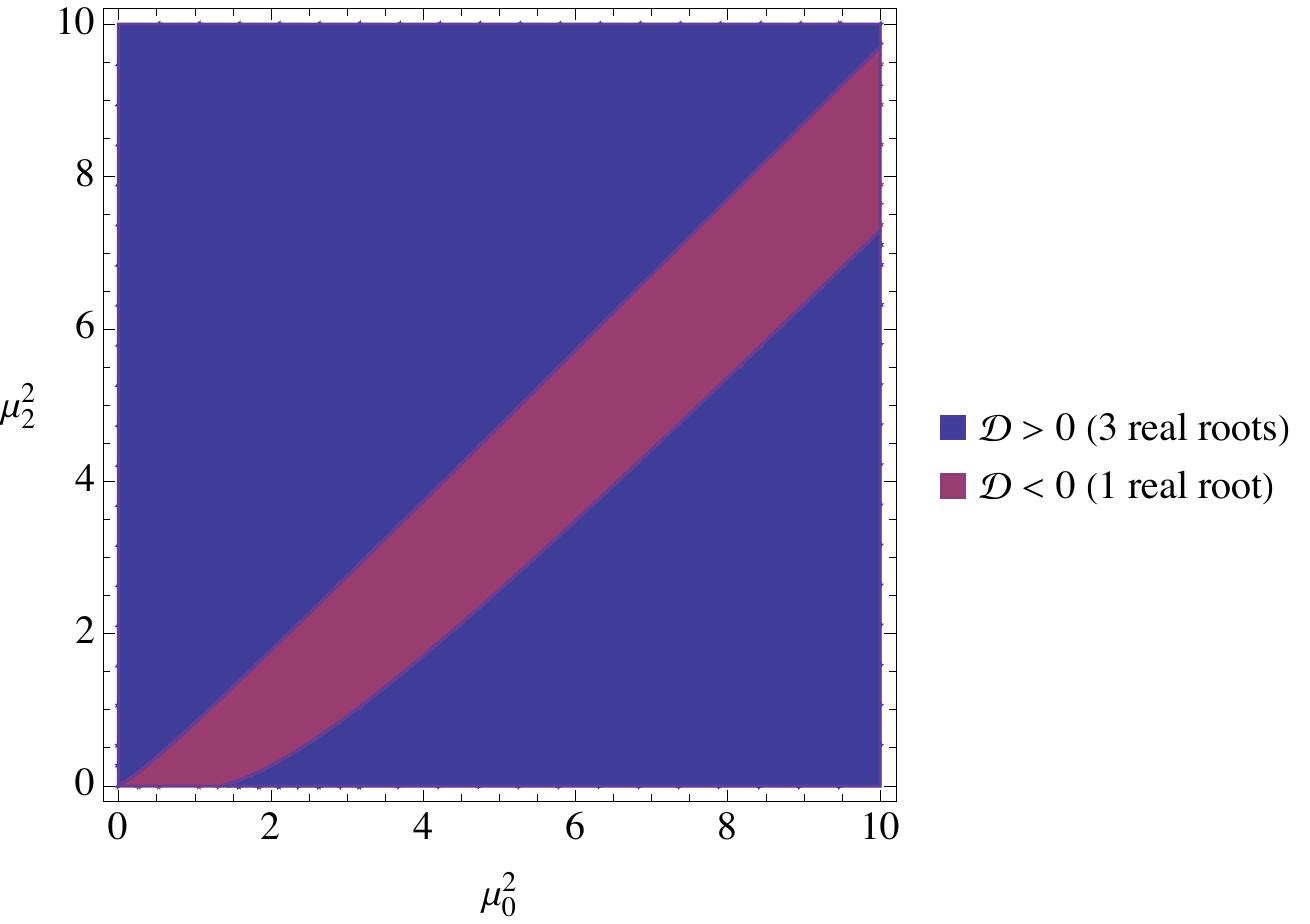}
\caption{\label{fig:disc}The sign of the discriminant $ \mathcal D $ \eqref{eq:disc}.}
\end{figure}

In the case of $ n = 1 $, the master equation \eqref{eq:master} becomes a linear homogeneous equation, which reads
\begin{equation}
f(\triangle_\xi)\,\theta
= 0
\label{eq:mastern=1}
\end{equation}
with the characteristic polynomial $ f $ being
\begin{equation}
f(x)
= x^3
  - (\mu_2^2 + \mu_0^2)\,x^2
  + (\mu_2^2\,\mu_0^2 + \alpha_2\,\mu_2^2 + \alpha_0\,\mu_0^2)\,x
  + \mu_2^2\,\mu_0^2\,.
\end{equation}
This problem can be treated in parallel to Sec.~\ref{sec:sol4n=1}, where we factorized the differential operator into a product of two Helmholtz operators.
Here, the sixth-order differential operator $ f(\triangle_\xi) $ can be cast into a product of three Helmholtz operators with eigenvalues given by the three roots of the characteristic equation $ f(x) = 0 $, and these roots characterize the solution of \eqref{eq:mastern=1}.
Having that the inflection point of $ f(x) $ lies at $ x = (\mu_2^2+\mu_0^2)/3 > 0 $ and $ f(0) = \mu_2^2\,\mu_0^2 > 0 $, the cubic equation $ f(x) = 0 $ turns out to have one and only one negative real root, which we denote as $ x = -\lambda_1^2 $ with $ \lambda_1 $ being real.
Whether the other two roots are positive real or complex depends on the sign of the discriminant
\begin{equation}
\begin{aligned}
\mathcal D
&
\equiv
  (\mu_2^2+\mu_0^2)^2\,(\mu_2^2\,\mu_0^2+\alpha_2\,\mu_2^2+\alpha_0\,\mu_0^2)^2
  - 4\,(\mu_2^2\,\mu_0^2+\alpha_2\,\mu_2^2+\alpha_0\,\mu_0^2)^3
  - 4\,(\mu_2^2+\mu_0^2)^3\,\mu_2^2\,\mu_0^2
  - 27 \mu_2^4\,\mu_0^4 \\
& \quad
  - 18\,
    (\mu_2^2+\mu_0^2)\,
    (\mu_2^2\,\mu_0^2+\alpha_2\,\mu_2^2+\alpha_0\,\mu_0^2)\,
    \mu_2^2\,\mu_0^2\,.
\end{aligned}
\label{eq:disc}
\end{equation}
The sign of $ \mathcal D $ in the parameter plane is shown in Fig.~\ref{fig:disc}.
In terms of the area, having $ \mathcal D \geq 0 $ (blue) is more likely as it generally realizes in the presence of a large hierarchy between the graviton masses, that is, when $ \mu_0 \gg \mu_2 $ or $ \mu_0 \ll \mu_2 $\,.
In this case, less massive graviton is expected to dominate.
The region of $ \mathcal D < 0 $ (red) is only seen around (to the slight right of) the equality line $ \mu_0^2 = \mu_2^2 $\,, in which the massive gravitons are expected to compete with each other.
In either case, we denote the two remaining roots as $ x = \lambda_2^2\,, \lambda_3^2 $\,, which are positive real if $ \mathcal D \geq 0 $, or complex and conjugate of each other if $ \mathcal D < 0 $.
Using Vi\`ete's formula, we may write the roots as
\begin{equation}
\begin{aligned}
-\lambda_1^2
&
= \frac{\mu_0^2+\mu_2^2}{3}
  + \frac{2}{3}\,\sqrt P\,
    \cos\left[\frac{1}{3}\,\cos^{-1}\left(\frac{Q}{2 P\,\sqrt P}\right) - \frac{4\pi}{3}\right]\,, \\
\lambda_2^2
&
= \frac{\mu_0^2+\mu_2^2}{3}
  + \frac{2}{3}\,\sqrt P\,
    \cos\left[\frac{1}{3}\,\cos^{-1}\left(\frac{Q}{2 P\,\sqrt P}\right) - \frac{2\pi}{3}\right]\,, \\
\lambda_3^2
&
= \frac{\mu_0^2+\mu_2^2}{3}
  + \frac{2}{3}\,\sqrt P\,
    \cos\left[\frac{1}{3}\,\cos^{-1}\left(\frac{Q}{2 P\,\sqrt P}\right)\right]
\end{aligned}
\end{equation}
with
\begin{equation}
\begin{aligned}
P
&
\equiv
  \mu_0^4 + \mu_2^4 - \mu_0^2\,\mu_2^2 - 3\,\sum_s \alpha_s\,\mu_s^2\,,\\
Q
&
\equiv
  2\,(\mu_0^6+\mu_2^6) - 3 \mu_0^2\,\mu_2^2\,(\mu_0^2+\mu_2^2) - 9\,\sum_s \alpha_s\,\mu_s^4 - 18 \mu_0^2\,\mu_2^2\,.
\end{aligned}
\end{equation}
With the use of these roots, the master equation settles down to the form
\begin{equation}
(\triangle_\xi + \lambda_1^2)\,(\triangle_\xi - \lambda_2^2)\,(\triangle_\xi - \lambda_3^2)\,\theta
= 0\,,
\end{equation}
which is ready to solve.

For $ \mathcal D > 0 $, the general solution is written in terms of real-valued functions as
\begin{equation}
\theta
= A_1\,\frac{\sin \lambda_1\,\xi}{\xi}
  + B_1\,\frac{\cos \lambda_1\,\xi}{\xi}
  + A_2\,\frac{\sinh \lambda_2\,\xi}{\xi}
  + B_2\,\frac{\cosh \lambda_2\,\xi}{\xi}
  + A_3\,\frac{\sinh \lambda_3\,\xi}{\xi}
  + B_3\,\frac{\cosh \lambda_3\,\xi}{\xi}\,.
\end{equation}
In the current case, unlike when $ n = 0 $, contributions from the two massive gravitons do not simply separate, as the eigenvalues $ \lambda_i $ depend on both of $ \mu_0 $ and $ \mu_2 $\,.
By imposing three boundary conditions $ \theta_\mathrm c = 1\,, \theta'_\mathrm c = \theta'''_\mathrm c = 0 $, we determine four constants as $ B_1 = B_2 = B_3 = 0 $ and $ A_1 = (1 - A_2\,\lambda_2 - A_3\,\lambda_3)/\lambda_1 $\,, obtaining the reduced expression\footnote{We find this different from Eq.~(38) of Ref.~\cite{Chen:2001a}.}
\begin{equation}
\theta
= \frac{\sin\lambda_1\,\xi
        + A_2\,\left(\lambda_1\,\sinh \lambda_2\,\xi - \lambda_2\,\sin\lambda_1\,\xi\right)
        + A_3\,\left(\lambda_1\,\sinh \lambda_3\,\xi - \lambda_3\,\sin\lambda_1\,\xi\right)}
       {\lambda_1\,\xi}\,.
\label{eq:thetan=1}
\end{equation}
To determine the remaining constants $ A_2 $ and $ A_3 $\,, we follow the same scheme as we employed in the $ n = 1 $ fourth-order case as follows.
On the one hand, these constants appear in the yet unused second and fourth derivatives as
\begin{equation}
\begin{aligned}
\theta''_\mathrm c
&
= \frac{-\lambda_1^2
        + A_2\,\lambda_2\,(\lambda_1^2 + \lambda_2^2)
        + A_3\,\lambda_3\,(\lambda_1^2 + \lambda_3^2)}
       {3}\,, \\
\theta^{(4)}_\mathrm c
&
= \frac{\lambda_1^4
        + A_2\,\lambda_2\,(\lambda_2^4-\lambda_1^4)
        + A_3\,\lambda_3\,(\lambda_3^4 - \lambda_1^4)}
       {5}\,.
\end{aligned}
\end{equation}
Then the boundary conditions on these derivatives in \eqref{eq:bc} provide us with the linear relations between the constants and the stellar integrals $ \iota_s $\,:
\begin{equation}
\begin{aligned}
A_2
&
= \frac{\lambda_1^2\,\lambda_3^2
        + \sum_{s=0,2} \alpha_s\,\mu_s^2\,
          [1 + (\lambda_3^2 - \lambda_1^2 - \mu_s^2)\,\mu_s\,\iota_s]}
       {\lambda_2\,(\lambda_1^2 + \lambda_2^2)\,(\lambda_3^2 - \lambda_2^2)}\,, \\
A_3
&
= \frac{\lambda_1^2\,\lambda_2^2
        + \sum_{s=0,2} \alpha_s\,\mu_s^2\,
          [1 + (\lambda_2^2 - \lambda_1^2 - \mu_s^2)\,\mu_s\,\iota_s]}
       {\lambda_3\,(\lambda_1^2 + \lambda_3^2)\,(\lambda_2^2 - \lambda_3^2)}\,.
\end{aligned}
\label{eq:C2C3}
\end{equation}
On the other hand, $ \iota_s $ may be calculated by substituting the profile function \eqref{eq:thetan=1} into Eq.~\eqref{eq:iota}.
Thanks to the simpleness of $ n = 1 $ polytrope, these integrations can be analytically done, and we are allowed to express $ \iota_s $ analytically in a form linear in $ A_2 $ and $ A_3 $\,, which however we do not present here as their expressions are too messy and not illuminating.
Then, substituting them into $ \iota_s $'s in \eqref{eq:C2C3} and solving for the integration constants $ A_2 $ and $ A_3 $ just algebraically, we obtain their expressions that include $ \mu_s $ and $ \xi_R $ only.
As a result, we arrive at the final analytical expression of the profile function $ \theta $ parametrically depending on $ \mu_s $ and $ \xi_R $\,.

The case with $ \mathcal D < 0 $ can be analyzed in parallel, or by means of analytical continuation, so we do not redo the procedure here but only show the general solution.
In this case, denoting the two complex conjugate roots as $ x = (p+q\,\mathrm i)^2\,, (p-q\,\mathrm i)^2 $\,, the general solution in terms of real-valued functions is written down as
\begin{equation}
\theta
= A\,\frac{\sin \lambda_1\,\xi}{\xi}
  + B\,\frac{\cos \lambda_1\,\xi}{\xi}
  + C_1\,\frac{\sinh p\,\xi\,\sin q\,\xi}{\xi}
  + C_2\,\frac{\cosh p\,\xi\,\sin q\,\xi}{\xi}
  + C_3\,\frac{\sinh p\,\xi\,\cos q\,\xi}{\xi}
  + C_4\,\frac{\cosh p\,\xi\,\cos q\,\xi}{\xi}\,.
\end{equation}
Lastly, in the special case with $ \mathcal D = 0 $\,, where the remaining roots degenerate, $ \lambda_2 = \lambda_3 $\,, the general solution is
\begin{equation}
\theta
= A\,\frac{\sin \lambda_1\,\xi}{\xi}
  + B\,\frac{\cos \lambda_1\,\xi}{\xi}
  + C\,\frac{\sinh \lambda_2\,\xi}{\xi}
  + D\,\frac{\cosh \lambda_2\,\xi}{\xi}
  + \tilde C\,\sinh \lambda_2\,\xi
  + \tilde D\,\cosh \lambda_2\,\xi\,.
\end{equation}

\begin{figure}[htbp]
\includegraphics[scale=.8]{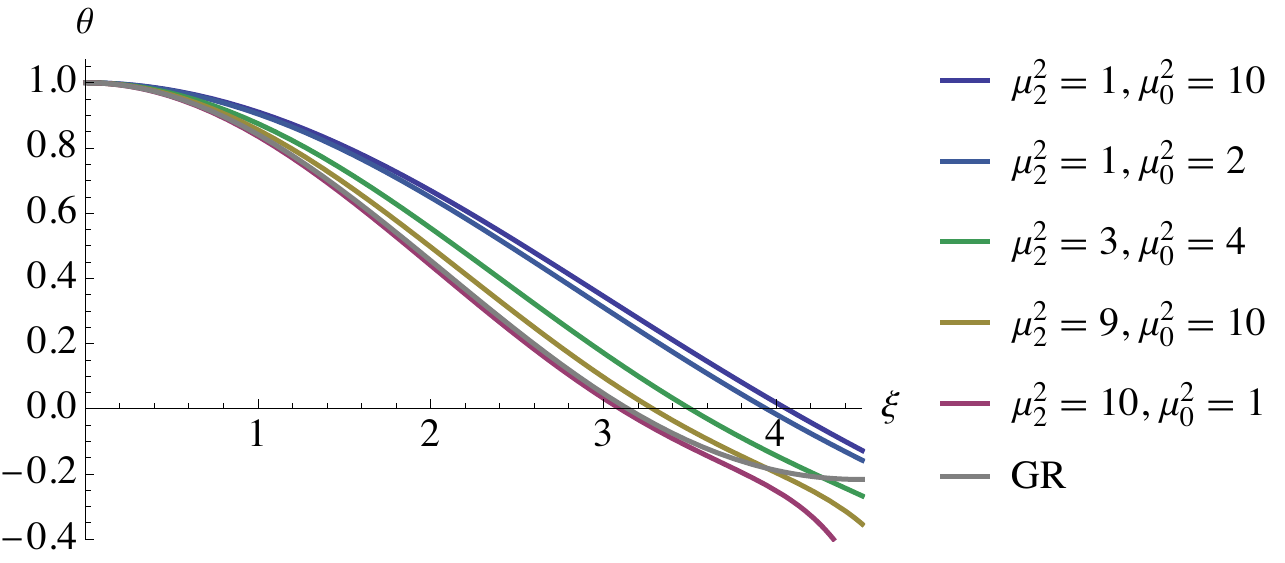}
\caption{\label{fig:soln=1}Typical solutions and the LE solution in GR (grey) for $ n = 1 $.}
\end{figure}

Figure~\ref{fig:soln=1} shows typical solutions for the polytropic index $ n = 1 $ together with the LE solution in GR (grey).
The tendencies can be well understood as consequences of the competition between the two massive gravitons, viz., attraction by the spin-$ 0 $ graviton dominates for $ \mu_0 \ll \mu_2 $ (red) while repulsion by the spin-$ 2 $ graviton dominates for $ \mu_2 \lesssim \mu_0 $ (rest).

\begin{figure}[htbp]
\includegraphics[scale=.8]{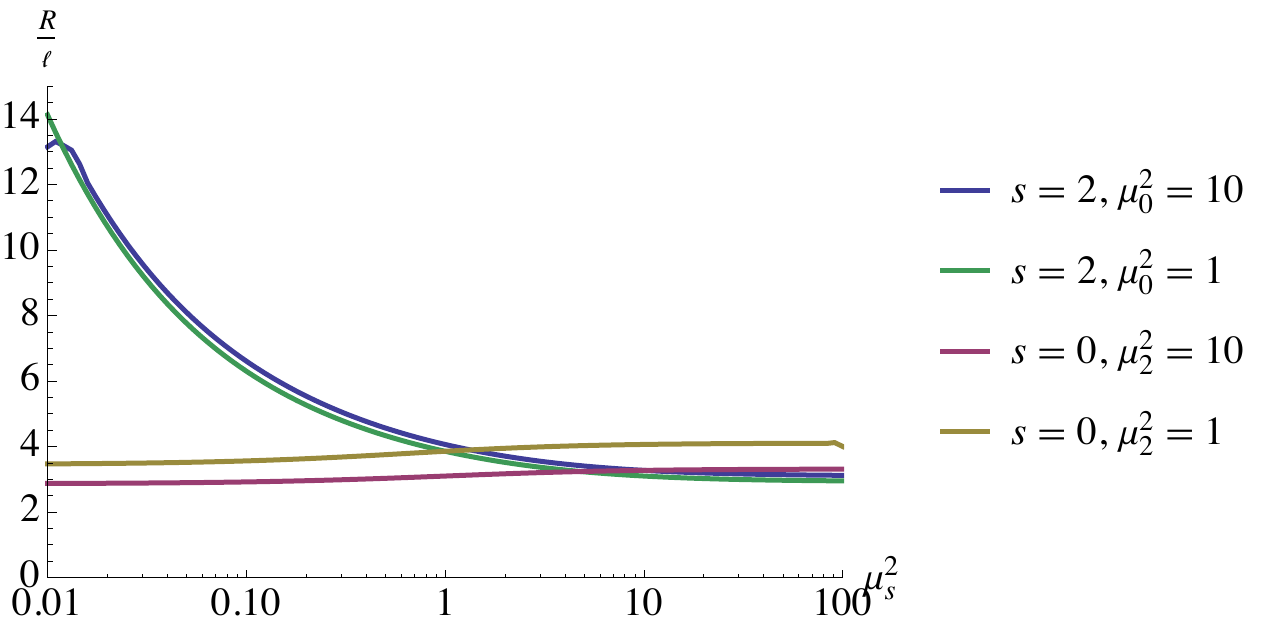}
\includegraphics[scale=.8]{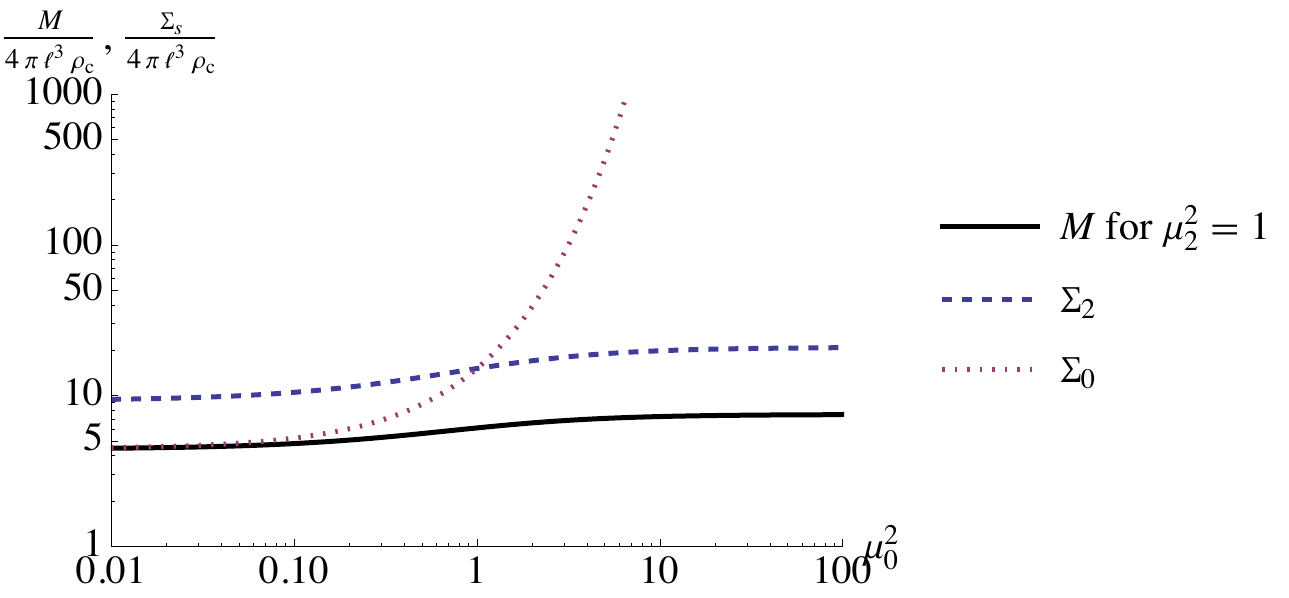}
\includegraphics[scale=.8]{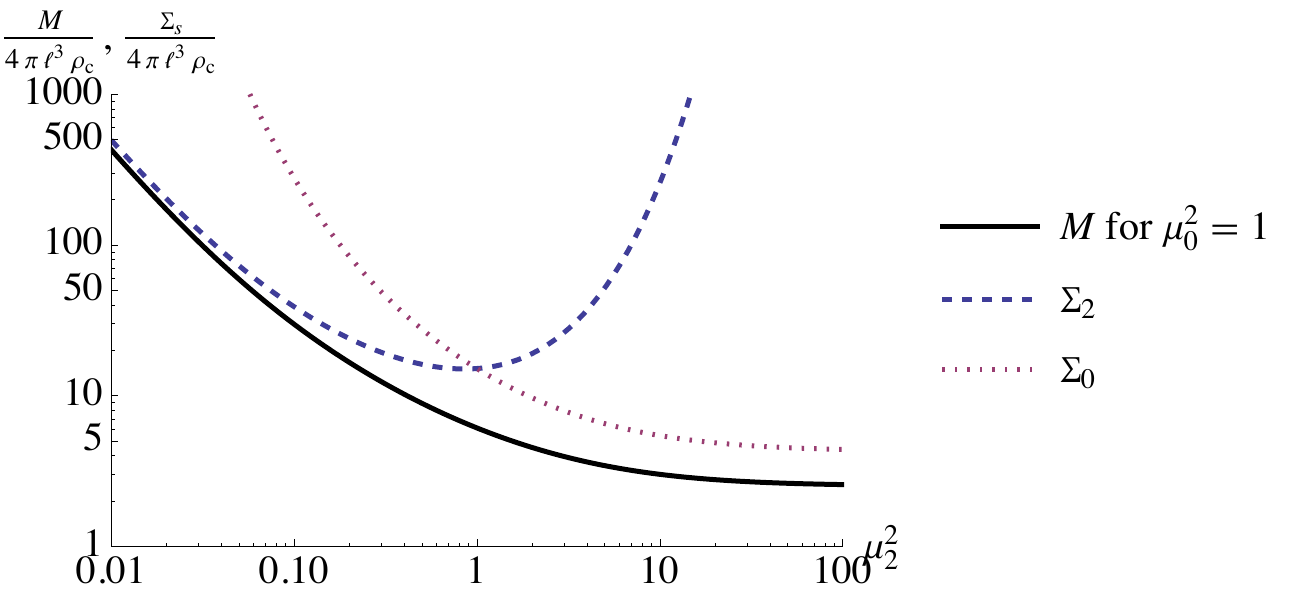}
\caption{\label{fig:RMSSn=1}Typical $ \mu_s $ dependences of $ R $ (top), $ M $, and $ \Sigma_s $ (middle and bottom) for $ n = 1 $. We believe the appearance of a kink near $ \mu_2^2 = 0 $ on the blue curve in the top panel is not physical but due to a lack of numerical precision in our calculation.}
\end{figure}

\begin{figure}[htbp]
\includegraphics[scale=.8]{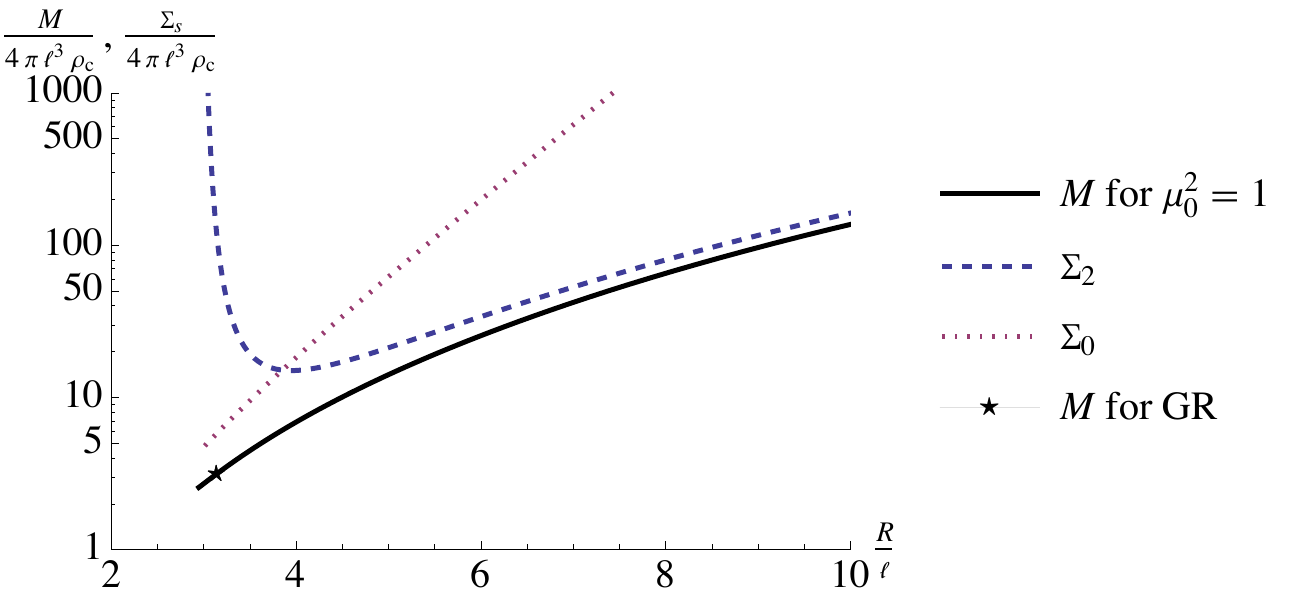}
\includegraphics[scale=.8]{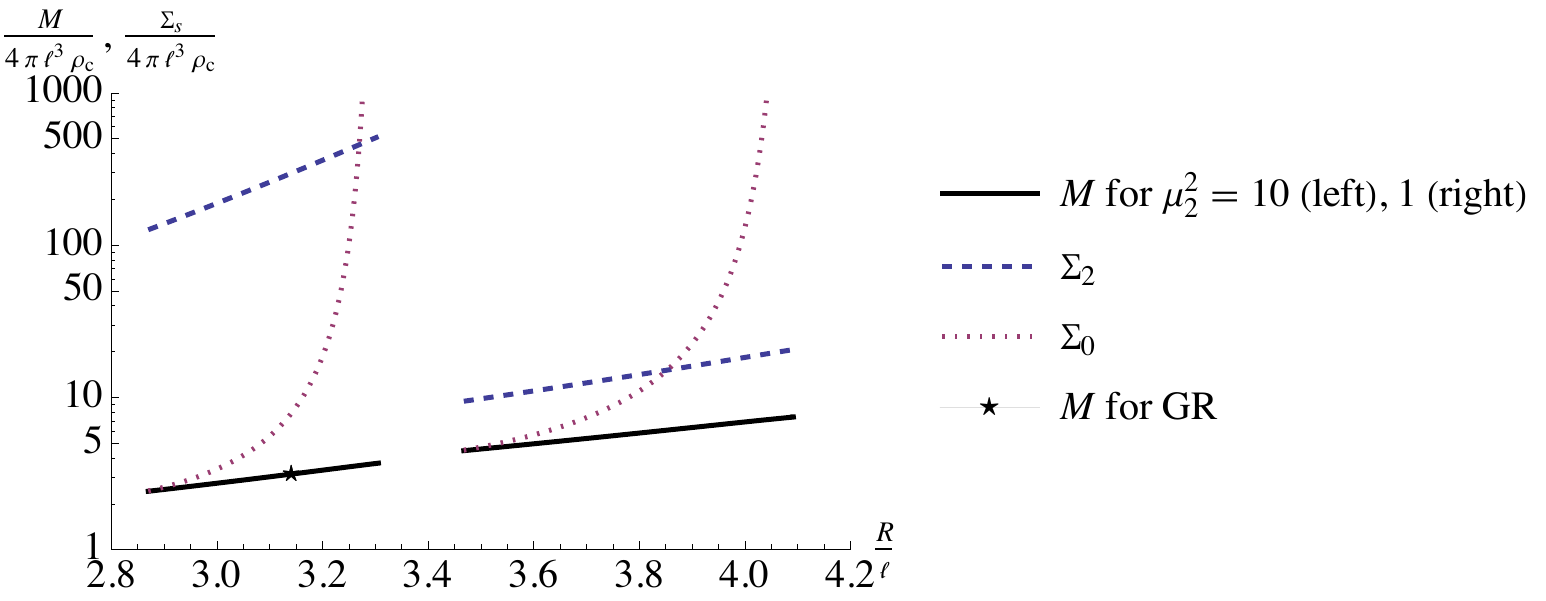}
\caption{\label{fig:MSS-Rn=1}Relationships between $ M $ and $ R $ (solid black) and $ \Sigma_s $ and $ R $ (dashed blue and dotted red) for $ n = 1 $.}
\end{figure}

Shown in Fig.~\ref{fig:RMSSn=1} are the $ \mu_s $ dependences of $ R $ (top), $ M $, and $ \Sigma_s $ (middle and bottom).
Figure~\ref{fig:MSS-Rn=1} shows the relationships between $ M $ and $ R $ (solid black) and $ \Sigma_s $ and $ R $ (dashed blue and dotted red).
All these appearances can be well understood in an analogous way to the previous cases.

\begin{figure}[htbp]
\includegraphics[scale=.8]{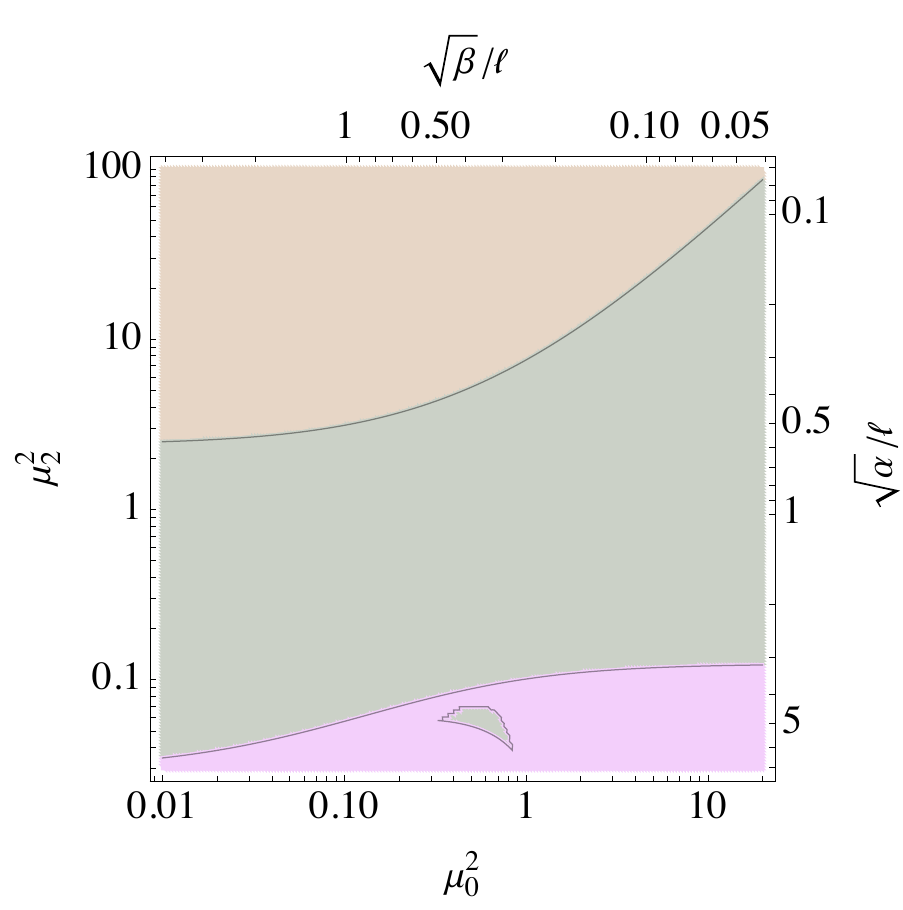}
\caption{\label{fig:ctn=1}Contours of the stellar radius $ R $ in the parameter plane for $ n = 1 $, on which the ratio $ R/R_\mathrm{LE} $ takes $ 1 $ (top) or $ 2 $ (bottom). We believe the appearance of a spot beneath the bottom contour is not physical but due to a lack of numerical precision in our calculation.}
\end{figure}

Two radius contours are plotted in Fig.~\ref{fig:ctn=1}, where on each curve the ratio $ R/R_\mathrm{LE} $ is $ 1 $ (top) and $ 2 $ (bottom).
From this diagram, one can conclude that the parameter $ \sqrt\alpha $ cannot exceed a few times $ \ell $ if one requires $ n = 1 $ polytrope stars have radius no larger than $ 2 $ or $ 3 $ times the GR value.

\section{\label{sec:concl}Conclusion}

In this paper, we studied non-relativistic polytropic stars in linearized higher-curvature theories of gravity (HCG).
Our particular aim was at formulating boundary conditions for the modified Lane--Emden equation with great care for the peculiarity arising from the massive nature of extra gravitons and providing a viable scheme for obtaining solutions to the boundary-value problems.

In Sec.~\ref{sec:le}, we analyzed the hydrostatic equilibrium condition, starting with the gauge-invariant equations of motion \eqref{eq:eom} derived from the second-order perturbative action \eqref{eq:action_s}.
In the static configuration, a particular set of gauge-invariant variables $ \Psi_2 $ and $ \Psi_0 $ as defined in \eqref{eq:Psi_s} turned to be useful to reduce the eoms into the decoupled form \eqref{eq:eom_Psi_s}.
These fourth-order eoms have the general solution in the form of difference of massless part $ \phi $ and massive part $ \psi_s $\,, which respectively satisfies the Poisson equation \eqref{eq:P} and the Helmholtz equation \eqref{eq:H}.
As a result, the gauge-invariant gravitational potential $ \Psi $, which appears in the hydrostatic equilibrium condition \eqref{eq:hydro}, was found as in \eqref{eq:Psi}.
The equilibrium condition is an integro-differential equation at this stage.
Applying an adequate higher-order differential operator on the both sides and adopting the polytropic equation of state, we obtained a sixth-order differential equation \eqref{eq:master} for the Lane--Emden-like variable $ \theta $.
When either of the graviton masses is taken to infinity, it reduces to a fourth-order equation \eqref{eq:master4} corresponding to ``$ R + R^2 $'' or ``$ R + C^2 $'' gravity.
When both go to infinity, it recovers the second-order Lane--Emden equation \eqref{eq:LE} in GR.

In Sec.~\ref{sec:bc}, we formulated the boundary conditions by focusing on the behavior of the potential at the stellar center.
This is because the derivatives of the profile function $ \theta $ are almost equivalent to those of the potential $ \Psi $ via \eqref{eq:hydro2}.
In order to have a necessary and sufficient number of boundary data for solving the sixth-order equation \eqref{eq:master}, we wrote down the derivatives at the stellar center up to fifth order as in \eqref{eq:bc} besides the same conditions \eqref{eq:bc_LE} as in GR.
We have proven the second and fourth derivatives cannot be determined locally but are related to integrals $ \iota_s $ \eqref{eq:iota} of the as-yet-undetermined profile function over the stellar interior.
This does not only mean the boundary conditions differ from GR, but also the nature of the boundary value problem drastically changes.
We also provided the analogous conditions for the fourth-order equation for ``$ R + R^2 $'' or ``$ R + C^2 $'' theories as in \eqref{eq:bc4}, in which the theory-dependent coefficient $ \alpha_s $ comes into play.

In Sec.~\ref{sec:sol}, we demonstrated how our scheme for solving the modified LE equations with the appropriate boundary conditions works for the polytrope indices $ n = 0 $ and $ 1 $, where in all cases analytical solutions exist.
In these cases the procedure for determining integration constants becomes trivial ($ n = 0 $) or reduces to solving a linear algebraic equation ($ n = 1 $).

In Sec.~\ref{sec:fourth}, we solved the fourth-order equation \eqref{eq:master4} imposing \eqref{eq:bc4}.
As shown in Figs.~\ref{fig:sol4n=0} ($ n = 0 $) and \ref{fig:sol4n=1} ($ n = 1 $), the dimensionless radius of the star increases (decreases) compared to GR for ``$ R + C^2 $'' (``$ R + R^2 $'') gravity, reflecting the repulsive (attractive) nature of the massive graviton.
In all cases, as $ \mu_s^2 \to \infty $, these solutions recover the Lane--Emden profile in GR \eqref{eq:sol_LE}.
The massless limit can be understood as a GR-like theory with a ``renormalized'' Newton constant $ (1+\alpha_s)\,G $.
In ``$ R + R^2 $'' gravity, $ \alpha_0 = 1/3 $, it mimics GR with a larger Newton constant $ 4G/3 $, leading to shrinkage of the radius by a factor of $ \sqrt 3/2 $, while the same limit of ``$ R + C^2 $'' gravity, $ \alpha_2 = -4/3 $, is antigravity with negative Newton constant $ -G/3 $, leading to an infinite radius.
We have clarified how the stellar radius $ R $, mass $ M $, and charge $ \Sigma_s $ depend on the graviton mass $ \mu_s $ in Figs.~\ref{fig:RMS4n=0} ($ n = 0 $) and \ref{fig:R4n=1}--\ref{fig:MS4n=1} ($ n = 1 $).
Diagrams relating the mass $ M $ and the charge $ \Sigma_s $ to the radius $ R $ were obtained in Figs.~\ref{fig:MS-R4n=0} ($ n = 0 $) and \ref{fig:MS-R4n=1} ($ n = 1 $).
We argued that, in ``$ R+C^2 $'' gravity, upper limits on the parameter $ \alpha $ in the action \eqref{eq:action} can be obtained by requiring the stellar radius $ R $ should not exceed several multiples of the GR values $ R_\mathrm{LE} $\,, which generally leads to $ \sqrt\alpha \lesssim \text{a few} \times \ell $.

In Sec.~\ref{sec:sixth}, we solved the sixth-order equation \eqref{eq:master} in generic HCG with the boundary conditions \eqref{eq:bc}.
Most of the modification trends as compared to GR arise as a result of competition of the opposite contributions from the co-existing massive gravitons.
In particular, when the masses have a large hierarchy, $ \mu_2 \ll \mu_0 $ or $ \mu_0 \ll \mu_2 $\,, the graviton with smaller mass dominates.
Because the coefficient of the massive gravitational potential for spin-$ 2 $, $ \alpha_2 $, is four times as large in magnitude as that of spin-$ 0 $, $ \alpha_0 $, the contribution from the former is generally more prominent than the latter when the two graviton masses are at the same order.
Typical solutions were presented in Figs.~\ref{fig:soln=0} ($ n = 0 $) and \ref{fig:soln=1} ($ n = 1 $).
The dependences of $ R $, $ M $, and $ \Sigma_s $ on $ \mu_s $ were shown in Figs.~\ref{fig:Rn=0}--\ref{fig:MSSn=0} ($ n = 0 $) and \ref{fig:RMSSn=1} ($ n = 1 $).
$ M $--$ R $ and $ \Sigma_s $--$ R $ relations were shown in Figs.~\ref{fig:MSS-Rn=0} ($ n = 0 $) and \ref{fig:MSS-Rn=1} ($ n = 1 $).
The dependence of the stellar radius $ R $, in the units of the GR value $ R_\mathrm{LE} $\,, on the mass parameters $ (\mu_0^2,\mu_2^2) $ were illustrated in Figs.~\ref{fig:ctn=0} ($ n = 0 $) and \ref{fig:ctn=1} ($ n = 1 $).
These will be useful to find allowed regions for the QCG parameters $ (\alpha,\beta) $ once an upper bound on the stellar radius of polytrope stars is established.

Finally, let us give some prospects for future studies.
On the theoretical side, development of an additional numerical procedure for imposing the boundary conditions becomes necessary if one wishes to construct solutions for an arbitrary polytropic index $ n $.
For $ n \neq 0,1 $, since no analytical solution is known, one has to somehow numerically make derivatives at the stellar center and integral of a solution over the stellar radius match.
We plan to present a viable scheme for this in a forthcoming paper.
On the observational side, the observable characteristics such as $ M $--$ R $ and $ \Sigma_s $--$ R $ diagrams, as well as the radius contours in the parameter plane, should offer a way to test HCG through comparisons with the distribution of known stellar populations.
We also plan to come back to this issue in the near future.

\begin{acknowledgments}
The authors are grateful to Hideki Asada for encouragements.
This work was in part supported by JST SPRING, Grant Number JPMJSP2152 (TT).
\end{acknowledgments}

\appendix

\section{\label{sec:gauge}Gauge transformations and gauge-invariant variables}

A general metric perturbation $ h_{\mu\nu} $ about a Minkowski background can be decomposed into scalar, vector, and tensor variables as
\begin{equation}
h_{\mu\nu}\,\mathrm dx^\mu\,\mathrm dx^\nu
= -2 A\,\mathrm dt^2
  - 2\,(\partial_i B + B_i)\,\mathrm dt\,\mathrm dx^i
  + 2\,(\delta_{ij}\,C +\partial_i \partial_j E + \partial_{(i} E_{j)} + H_{ij})\,
    \mathrm dx^i\,\mathrm dx^j\,,
\end{equation}
where vector and tensor variables satisfy $ \partial_i B^i = \partial_i E^i = \partial_i H^{ij} = H_i{}^i = 0 $ and the parentheses around tensor indices denote symmetrization.
An active coordinate transformation $ x^\mu \rightarrow x^\mu + \xi^\mu(x) $ with $ \xi^\mu $ being as small as $ h_{\mu\nu} $ in magnitude transforms the metric perturbation, to first order, as
\begin{equation}
h_{\mu\nu}
\rightarrow
  h_{\mu\nu} - \pounds_\xi \eta_{\mu\nu}\,,
\end{equation}
where $ \pounds_\xi $ is the Lie derivative along $ \xi^\mu $\,.
$ \xi^\mu $ can be decomposed into the scalar and vector parts as $ (\xi^\mu) = (T,\partial^i L + L^i) $ with $ \partial_i L^i = 0 $.
It is obvious that this does not affect the tensor variable:
\begin{equation}
H_{ij}
\rightarrow
  H_{ij}\,.
\end{equation}
On the other hand, the vector variables are transformed as
\begin{equation}
B_i
\rightarrow
  B_i + \dot L_i\,,
\quad
E_i
\rightarrow
  E_i - L_i\,,
\end{equation}
where the dot denotes differentiation with respect to $ t $.
Hence, the following combination is found to be invariant:
\begin{equation}
\Sigma_i
\equiv
  B_i + \dot E_i\,.
\end{equation}
The transformations of the scalar variables are
\begin{equation}
A
\rightarrow A
  -\dot T\,,
\quad
B
\rightarrow
  B - T + \dot L\,,
\quad
C
\rightarrow
  C\,,
\quad
E
\rightarrow
  E - L\,,
\end{equation}
from which a useful set of invariant combinations is found to be
\begin{equation}
\Psi
\equiv
  A - \dot B - \ddot E\,,
\quad
\Phi
\equiv
  C\,.
\end{equation}

\section{\label{sec:pert}Gauge-invariant expressions for the higher-curvature Lagrangians}

In terms of the gauge-invariant variables, the linear perturbation of the Ricci tensor and Ricci scalar on a Minkowski background are written as
\begin{equation}
\begin{aligned}
{}^{(1)}R_{ij}
&
= -\square H_{ij}
  + \partial_{(i} \dot\Sigma_{j)}
  - \partial_i \partial_j \Psi
  - \partial_i \partial_j \Phi
  - \delta_{ij}\,\square \Phi\,, \\
{}^{(1)}R_{i0}
&
= \frac{1}{2}\,\triangle \Sigma_i
  - 2 \partial_i \dot\Phi\,, \\
{}^{(1)}R_{00}
&
= \triangle \Psi
  - 3 \ddot\Phi
\end{aligned}
\end{equation}
and
\begin{equation}
{}^{(1)}R
= -2 \triangle(\Psi-\Phi)
  - 6 \square\Phi\,,
\end{equation}
respectively.

The topological nature of the Gauss--Bonnet combination in four dimensions allows us to rewrite the Weyl-squared action, up to irrelevant surface integrals, as
\begin{equation}
S_C
\equiv
  \frac{-\alpha}{16\pi\,G}\,
  \int\!\mathrm d^4x\,\sqrt{-g}\,C_{\mu\nu\rho\sigma}\,C^{\mu\nu\rho\sigma}
= \frac{-\alpha}{16\pi\,G}\,
  \int\!\mathrm d^4x\,\sqrt{-g}\,\left(2 R_{\mu\nu}\,R^{\mu\nu} - \frac{2}{3}\,R^2\right)\,.
\end{equation}
This is then expanded up to second order in the perturbative variables as
\begin{equation}
\begin{aligned}
{}^{(2)}S_C
&
= \frac{-\alpha}{16\pi\,G}\,\int\!\mathrm d^4x\,\left(
   2 {}^{(1)}R_{\mu\nu}\,{}^{(1)}R^{\mu\nu}
   - \frac{2}{3}\,{}^{(1)}R^2
  \right) \\
&
= \frac{-\alpha}{16\pi\,G}\,\int\!\mathrm d^4x\,
  \left[
   2 (\square H_{ij})^2
   + (\partial_i \dot\Sigma_j)^2
   - (\triangle\Sigma_i)^2
   + \frac{4}{3}\,\left[\triangle (\Psi-\Phi)\right]^2
  \right]\,,
\end{aligned}
\end{equation}
where surface terms have been discarded.
The second-order perturbation of the Ricci-squared action
\begin{equation}
S_R
\equiv
  \frac{\beta}{16\pi\,G}\,\int\!\mathrm d^4x\,\sqrt{-g}\,R^2
\end{equation}
is
\begin{equation}
\begin{aligned}
{}^{(2)}S_R
&
= \frac{\beta}{16\pi\,G}\,\int\!\mathrm d^4x\,{}^{(1)}R^2 \\
&
= \frac{\beta}{16\pi\,G}\,\int\!\mathrm d^4x\,
  4 \left[\triangle (\Psi-\Phi) + 3 \square\Phi\right]^2\,.
\end{aligned}
\end{equation}

The interaction Lagrangian for a perturbative energy-momentum tensor $ T_{\mu\nu} $ minimally coupled to gravity is
\begin{equation}
S_\mathrm{int}
= \frac{1}{2}\,\int\!\mathrm d^4x\,h_{\mu\nu}\,T^{\mu\nu}\,,
\end{equation}
where $ T_{\mu\nu} $ can be decomposed into scalar, vector, and tensor variables as
\begin{equation}
T_{00}
= \epsilon\,,
\quad
T_{0i}
= -\partial_i v - v_i\,,
\quad
T_{ij}
= P\,\delta_{ij}
  + \left(\partial_i \partial_j - \frac{1}{3}\,\delta_{ij}\,\triangle\right)\,\sigma
  + \partial_{(i} \sigma_{j)}
  + \sigma_{ij}\,.
\end{equation}
We assume the conservation law $ \partial_\mu T^{\mu\nu} = 0 $ holds, which settles down in the decomposed form:
\begin{equation}
\dot\epsilon + \triangle v = 0\,,
\quad
\dot v + P + \frac{2}{3}\,\triangle\sigma = 0\,,
\quad
\dot v^i + \frac{1}{2}\,\triangle \sigma^i = 0\,.
\end{equation}
Then the interaction Lagrangian is rewritten in terms of the gauge-invariant variables as
\begin{equation}
\begin{aligned}
S_\mathrm{int}
&
= \int\!\mathrm d^4x\,\left[
   -A\,\epsilon
   + B\,\triangle v
   + 3 C\,P
   + E\,\triangle\,\left(P + \frac{2}{3} \triangle\,\sigma\right)
   - B_i\,v^i
   - \frac{1}{2}\,E_i\,\triangle \sigma^i
   + H_{ij}\,\sigma^{ij}
  \right] \\
&
= \int\!\mathrm d^4x\,\left[
   -\Psi\,\epsilon
   + 3 \Phi\,P
   - \Sigma_i\,v^i
   + H_{ij}\,\sigma^{ij}
  \right]\,,
\end{aligned}
\end{equation}
where surface terms have been discarded.

\section{\label{sec:Helm}Solution for higher-order Helmholtz equations}

We consider a linear inhomogeneous equation of the form
\begin{equation}
f(\triangle)\,\varphi
= S\,,
\end{equation}
where $ f $ is an $ n $-th order polynomial, which we call the characteristic function, $ \triangle $ the flat-space Laplace operator, and $ S $ a given source function.
Without loss of generality, using the $ n $ roots for the characteristic equation $ f(x) = 0 $, $ x = \lambda_i $ ($ i = 1,\cdots,n $), the problem reduces to solving
\begin{equation}
\prod_{i=1}^n \left(\triangle - \lambda_i\right)\,\varphi
= S\,.
\end{equation}
We assume $ \lambda_i \neq \lambda_j $ for $ i \neq j $ for simplicity, but extending the formula to degenerate cases is straightforward.
The above equation admits a solution of the form $ \varphi = \sum_{i=1}^n \varphi_i $\,, where each $ \varphi_i $ solves a single Helmholtz equation
\begin{equation}
\left(\triangle - \lambda_i\right)\,\varphi_i
= c_i\,S
\end{equation}
with the coefficients $ c_i $ ($ i = 1,\cdots, n $) satisfying the following system of $ n $ linear equations
\begin{equation}
\begin{aligned}
&
\sum_{i=1}^n c_i = 0\,, \\
&
\sum_{i=2}^n c_i\,(\lambda_i - \lambda_1) = 0\,, \\
&
\sum_{i=3}^n c_i\,(\lambda_i - \lambda_1)\,(\lambda_i - \lambda_2) = 0\,, \\
&
\vdots \\
&
\sum_{i=k}^n c_i\,\prod_{j=1}^{k-1} (\lambda_i - \lambda_j) = 0\,, \\
&
\vdots \\
&
\sum_{i=n-1}^n c_i\,\prod_{j=1}^{n-2} (\lambda_i - \lambda_j) = 0\,, \\
&
c_n\,\prod_{j=1}^{n-1} (\lambda_n - \lambda_j) = 1\,.
\end{aligned}
\end{equation}

\bibliography{hcgle}

\end{document}